\documentclass[sigplan, 9pt]{acmart}

\usepackage{booktabs} 

\usepackage[noend]{algpseudocode}
\usepackage{algorithm}

\usepackage{amsmath}
\usepackage{diagbox}
\usepackage{subcaption} 

\usepackage{tabulary}
\usepackage{tabularx}
\usepackage{makecell}
\usepackage[textsize=scriptsize]{todonotes}

\usepackage{pgfplots}
\usepackage{tikz}
\usetikzlibrary{
arrows.meta,
colorbrewer,
decorations.pathreplacing }

\usepackage{balance}

\usepackage{fancyhdr}

\acmYear{2019}\copyrightyear{2019}
\setcopyright{acmcopyright}
\acmConference[Middleware '19]{Middleware '19: Middleware '19: 20th International Middleware Conference}{December 8--13, 2019}{Davis, CA, USA}
\acmBooktitle{Middleware '19: Middleware '19: 20th International Middleware Conference, December 8--13, 2019, Davis, CA, USA}
\acmPrice{15.00}
\acmDOI{10.1145/3361525.3361548}
\acmISBN{978-1-4503-7009-7/19/12}

\newcommand{\framework}{eSPICE}

\usepackage[most]{tcolorbox}
\fancypagestyle{firstpage}{%
	\rhead{}
	\lhead{}
	\cfoot{	
	\begin{tcolorbox}[colback=white,
		colframe=red,
		width=13cm,
		arc=3mm, auto outer arc,
		]
		{\color{red}
		\copyright ~ACM, [2019]. This is the author's version of the work. It is posted here by permission of ACM for your personal use. Not for redistribution. The definitive version  will be published in the Proceedings of ACM Middleware 2019.}
	\end{tcolorbox}
	}
}

\begin{document}
\title{eSPICE: Probabilistic Load Shedding from Input Event Streams in Complex Event Processing}
\renewcommand{\shorttitle}{eSPICE}


\author{Ahmad Slo, Sukanya Bhowmik,  Kurt Rothermel}
\orcid{1234-5678-9012}
\affiliation{%
  \institution{University of Stuttgart}
}
\email{firstName.lastName@ipvs.uni-stuttgart.de}
%

\begin{abstract}
Complex event processing systems process the input event streams on-the-fly. Since input event rate could overshoot the system's capabilities and results in violating a defined latency bound, load shedding is used to drop a portion of the input event streams. The crucial question here is how many and which events to drop so the defined latency bound is maintained and the degradation in the quality of results is minimized. In stream processing domain, different load shedding strategies have been proposed but they mainly depend on the importance of individual tuples (events). However, as complex event processing systems perform pattern detection, the importance  of events is also influenced by other events in the same pattern. In this paper, we propose a load shedding framework called \framework~for complex event processing systems. \framework~depends on building a probabilistic model that learns about the importance of events in a window. The position of an event in a window and its type are used as features to build the model. Further, we provide algorithms to decide when to start dropping events and how many events to drop. Moreover, we extensively evaluate the performance of \framework~on two real-world datasets.

\end{abstract}

%
%

 

\settopmatter{printfolios=true} 

\maketitle
\thispagestyle{firstpage}

\section{Introduction}
Complex event processing (CEP) is an efficient and scalable paradigm for processing event streams. An operator in a CEP system is used to detect important situations by analyzing the input event streams. It performs pattern matching by correlating the events (called primitive events) from the input event streams and generates, as output, complex events which represent the occurrence of specific situations, e.g., fire detection, stock changes, intrusion detection, etc. \cite{1:cu2012processing, spectre:2017, 3:Olston:2003:AFC:872757.872825}. In CEP  \cite{1:Balkesen:2013:RRI:2488222.2488257, CastroFernandez:2013:ISO:2463676.2465282, 1:flux:1260779, 1:s4:5693297, 1:apache_storm}, the input event stream is partitioned into independent windows of events where a window captures temporal relations between events. Windows might overlap and hence an event can be part of several windows.

In CEP, the volume of input event streams is huge and cannot be processed on a single machine. Therefore, distribution and parallelism are frequently used  in CEP, where the CEP operator graph is distributed on multiple compute nodes. Moreover, each CEP operator runs on one or more compute nodes \cite{CastroFernandez:2013:ISO:2463676.2465282, 1:s4:5693297, spectre:2017, 1:Balkesen:2013:RRI:2488222.2488257, skipping:lima}.  The underlying assumption of the above works is that there are infinite available resources, e.g., in cloud.  
However, there are various reasons for considering limited resources such as: 1) limited monetary budget, and 2) limited compute resources if operators run in private clouds due to security or response time reasons.

In most CEP applications, the detected complex events are useless if they are not detected within a certain latency bound \cite{Quoc:2017:SAC:3135974.3135989, CastroFernandez:2013:ISO:2463676.2465282}. Moreover, many applications accept a reduced result quality, e.g., network monitoring, traffic monitoring, stock market \cite{3:Olston:2003:AFC:872757.872825, 1:Balkesen:2013:RRI:2488222.2488257, Zacheilas:2015}. 
To avoid violating a defined latency bound or crashing the system in the face of high incoming event rate and limited resources, load shedding may be necessary. Load shedding drops events from input event streams of an operator, thereby reducing its load. However, dropping events might adversely impact the quality of the CEP output where important situations could be missed or falsely detected in the input event stream. Thus, it is crucial to drop those events that have less impact (low utility/importance) on the quality of results. 

There are primarily three challenges facing the decision to drop events in CEP systems: 
1) Deciding on which events to drop since the utility of an event depends on multiple factors, e.g., other events in the pattern, on the order of events in the pattern, and on the input event stream.
2) Calculating the number of events to drop in order to maintain a given latency bound since an event may be dropped from some windows while it is still there in other windows.  
3) Dropping events in an efficient way to reduce the overhead of load shedding.

Load shedding has been proposed by several research groups \cite{3:Tatbul:2003:LSD:1315451.1315479, 3:Tatbul:2006:WLS:1182635.1164196, 3:Rivetti:2016:LSS:2933267.2933311, 3:Olston:2003:AFC:872757.872825, 3:Kalyvianaki:2016:TFF:2882903.2882943, 8622265} in the stream processing domain. They focus mainly on individual events, where each event has an associated utility value that reflects its importance for the result quality.  
However, since CEP systems perform pattern matching, the utility of events is also influenced by other events in the same pattern. For example, let the pattern be $\{seq(A;B) \lor seq (A;C)\}$. It is clear that events of type $A$ are more important than events of type $B$ and $C$. However, if all events of type $B$ and $C$ are dropped, then no complex events would be detected. Hence, we cannot consider only the utility of each event individually but we must also take into consideration other events in the pattern and in the input event streams. 
So far, there is only little work on load shedding in CEP. In \cite{3:He2014OnLS}, the authors consider the dependency between different events of the same pattern. However, they do not consider the order of events in patterns and input event streams which is extremely important in CEP such as in the sequence operator. 

In this paper, we propose a load shedding framework, called \framework, for CEP systems. \framework~is efficient and lightweight. Moreover, it considers the dependency between events of the same pattern as well as the order of events in the pattern and in input event streams. 
In addition, it also considers the impact of the same event residing in overlapping windows on the quality of results, where the same event may be in different positions in different windows. To capture the utility of an event in different windows, we design a probabilistic model that uses the \textbf{relative position} of events in a window and their \textbf{types} as learning features. 
The goal of our load shedding strategy is to maintain a given latency bound while minimizing the adverse impact of dropping events on the quality of results. 
More specifically, our contributions in this paper are as follows: 
\begin{itemize}
	\item   We propose an efficient lightweight load shedding strategy that uses a probabilistic model to capture the importance of events in a window. The importance of an event is influenced by its type and its relative position in a window. The idea behind this approach is that the events, in specific positions within a window, that contribute to building a complex event in one window are more likely to build complex events in other windows as well.
	
	\item We provide an algorithm to estimate the number of events to drop in order to maintain the given latency bound. It also estimates the intervals within which the drop should take place.
	
	\item In order to show the effectiveness of our proposed load shedding strategy under realistic settings, we implement and thoroughly evaluate \framework~for a broad range of CEP operators using real world datasets.
	
\end{itemize}

The rest of the paper is structured as follows. Section 2 presents the background for this work and formally states the problem that we address. In Section 3, we provide a detailed explanation of how the different components of \framework~are used for our probabilistic load shedding strategy.  Section 4 presents results obtained from extensively evaluating \framework. Finally,  we discuss the related work and the conclusion in Sections 5 and 6, respectively.

\section{Preliminaries and Problem Statement}
\label{sec:background}
\label{sec:systemModel}
In complex event processing (CEP) systems, input event streams are processed to detect patterns. Such CEP systems may comprise of one or more operators that are represented by a directed acyclic graph.
Each operator processes input event streams produced from one or more sources. Event sources might be sensors, upstream operators, other applications, etc. An event (also called primitive event) in the input event streams consists of a meta-data and attribute-value pairs. The event meta-data contains event type, sequence number and timestamp. The event type is used to distinguish events from different types. The event type could be, for example, a stock symbol in a stock market application, a player in a soccer application, etc. 
Events in the input event streams have global order, e.g., by using the sequence number or the timestamp and a tie-breaker. The attribute-value pairs contain the actual data, e.g., stock quote or player position.  
      

In this work, we focus on a CEP system consisting of a single CEP operator. We assume a window-based CEP system where the input event streams are partitioned into windows (for example, by a window operator) using predicates. The predicates to open and close windows may depend on time (called time-based window), on the number of events (called count-based window), on logical predicates (called pattern-based window) or, on a combination of them \cite{1:Balkesen:2013:RRI:2488222.2488257, Mayer:2017:MCO:3093742.3093914}.
The windows of primitive events are first pushed to the input queue of an operator. 
An operator continuously gets primitive events from the input queue and processes them  by the \textit{process} function which performs the actual pattern matching as shown in Figure \ref{fig:ls}. The output of the pattern matching are the \textit{complex events}.  A primitive event might belong to multiple overlapped windows where it is processed \textit{independently} in each window. 


To define patterns, the operator uses an event specification language like Tesla \cite{1:cu2010tesla}, Snoop \cite{1:ch1994snoop}, or SASE \cite{Wu:2006:HCE:1142473.1142520}. These languages contain several CEP operators: sequence, conjunction, negation, etc.

\textbf{Example:} In intra-day stock trading, an operator receives an event stream containing live stock quote changes of stock $A$ and $B$ throughout the trading day. An analyst wants to detect correlations between a change in $A$ and a change in $B$ in a time period of 1 minute. He formulates a query in the Tesla \cite{1:cu2010tesla} event specification language:
\begin{equation*}[Q_E]
	\begin{aligned}
		\ & \mathtt{define\ Influence(Factor)} \\
		\ & \mathtt{from\  B()\ and } \\
		\ & \mathtt{A()\ within\ 1min\ from\ B }\\
		\ & \mathtt{where\ Factor = B:change\ /\ A:change }
	\end{aligned}
\end{equation*}

Let us assume that a window $w$ of 1 minute contains the stock events  $B_4$, $B_3$, $A_2$, $A_1$, where $X_i$ represents event type $X$ and $i$ indicates the event position in stream of ordered events. There exist several instance of each event type in $w$ and hence it is not clear which events of type $A$ should be matched with which events of type $B$. The generated complex events could be any combination of these event instances. To precisely define which event instances should participate in emitting the complex events, the \textit{selection policy} has been introduced \cite{1:ch1994snoop, 1:Zimmer:1999:SCE:846218.847253}. There are four main selection policies: \textit{first}, \textit{last}, \textit{each}, \textit{cumulative}. In the first selection policy, the earliest event instances are chosen for pattern matching.  In the above example, the generated complex events using the \textit{first} selection policy could be $cplx_{13}=$ ($A_1$, $B_3$) and  $cplx_{24}=$ ($A_2$, $B_4$). In last selection policy, the latest event instances are chosen for pattern matching. In the above example, the generated complex events using the \textit{last} selection policy could be $cplx_{23}=$ ($A_2$, $B_3$) and $cplx_{24}=$ ($A_2$, $B_4$).

In the above example, it is also not clear whether it is allowed to reuse the event instances in performing other pattern matching or they should be considered as consumed events, i.e., not reusing them again. The \textit{consumption policy} \cite{1:ch1994snoop, 1:Zimmer:1999:SCE:846218.847253} defines whether the same event instance can be used in several pattern matching. There are mainly two consumption policies: \textit{consumed} and \textit{zero}. In the above example again, let us assume that the selection policy is \textit{last}. Now, using \textit{consumed} consumption policy results in detecting only one complex event $cplx_{23}=$ ($A_2$, $B_3$). Whilst using \textit{zero} consumption policy results in detecting two complex events $cplx_{23}=$ ($A_2$, $B_3$) and $cplx_{24}=$ ($A_2$, $B_4$), where the event $A_2$ is reused in $cplx_{24}$. For more information about selection and consumption policy, see \cite{1:ch1994snoop, 1:Zimmer:1999:SCE:846218.847253}.

Depending on the used selection and consumption policy, the events in specific positions within a window have different probabilities to be part of the detected complex events. For example, if the used selection policy is \textit{first} and consumption policy is \textit{zero}, then the events in the beginning of the windows have higher probabilities to match the pattern.
  
We assume that the processing logic in the operator is a black-box, where we do not have any knowledge about their internal states. In this work, we assume that the operator reveals detected complex events, which is a standard assumption in any event processing system. In addition, we assume that the event types are known.
Our probabilistic model estimates the utility of events regardless of their order within the matching patterns and also regardless of the used selection and consumption policy.


\subsection{Quality of Results}
In this paper, we represent the quality of results by number of false positives and negatives. A false positive is a situation (a complex event) that should not be detected but has been falsely detected. While a false negative is a situation (a complex event) that should be detected but has not been detected. 

In the above example (cf. \ref{sec:systemModel}), using the  \textit{first} selection policy and \textit{consumed} consumption policy, two complex events can be detected $cplx_{13}=$ ($A_1$, $B_3$) and    $cplx_{24}=$ ($A_2$, $B_4$). If $A_2$ is dropped from  $w$, only one complex event can be detected $cplx_{13}=$ ($A_1$, $B_3$). This results in one false negative since $cplx_{24}$ is not detected. On the other hand, if $A_1$ is dropped  from $w$, a \textit{new} complex event is detected $cplx_{23}=$ ($A_2$, $B_3$) which results in one false positive and two false negatives since $cplx_{13}$  and $cplx_{24}$ are not detected.

\subsection{Problem Statement}
To maintain a given latency bound ($LB$) in the face of system overload, load shedding could be used to drop events from the operator's input queue.  But dropping events might degrade the quality of result. To minimize its impact on the quality of result, the load shedder must drop only those events that have low utility values.
The utility of primitive events is measured by their influence on the number of false positives ($N_{fp}$) and false negatives ($N_{fn}$), i.e., the quality.

More formally, our objective is to minimize degradation in the quality of result, i.e., minimize ($N_{fp} + N_{fn}$), while dropping events such that the given latency bound $LB$ is met.

\section{Probabilistic Load Shedding}
\label{sec:ls}
In order to minimize the degradation in quality of results, our main idea is to avoid dropping primitive events that could contribute to producing complex events. The question is--how do we identify the importance/utility, in this context, of these primitive events prior to processing them? 
In real-world applications, event streams have properties that can be exploited to derive the aforementioned importance/utility of a primitive event w.r.t. its probability of contributing to a complex event. An observation is that there is a correlation between \textbf{type} and \textbf{relative position} within windows of primitive events that constitute complex events. 
For example, in a soccer game, a sport analyst might be interested in finding a complex event called \textit{man- marking}, i.e., certain defender(s) who always defend against a particular striker. In this case, the ball possession by a striker (\textit{possession-event}) and the defender (\textit{defending-event}) are event types. These two event types have a correlation with each other. Whenever a striker possesses the ball, a defender(s) defends against him in a certain time interval (i.e., relative position in the window), thus producing a complex event.   
Clearly, in this scenario, position of the primitive events constituting the complex events, i.e., position of \textit{defending-events} relative to the \textit{possession-event}, are correlated. 
Such correlations also exist in stock market applications. For example, a stock of type IBM may impact a stock of another company within a certain time interval (i.e., relative position in the window), thus resulting in a complex event which detects such an influence. Again, in a different domain,  the sensor  data set provided by the Intel Research Berkeley Lab shows positive correlation between events of type temperature and events of type humidity \cite{8365752}. This implies that within a certain interval an increase/decrease in temperature results in an increase/decrease in humidity.

We exploit this correlation, captured by the type and relative position in the window of primitive events, to predict the probability of primitive events to be part of a complex event. In particular, we derive utilities of primitive events in a \textbf{window} based on the event \textbf{types} and their \textbf{relative positions} within the window and use this information to build a probabilistic model that estimates the utility of incoming events in windows.
Our load shedder drops only the incoming events that have low utility values within each window, thus minimizing the number of false positives and false negatives.

Next, we explain our probabilistic load shedding strategy in detail. First, we show the architecture of \framework. Then, we formally define the utility of primitive events within windows. This is followed by a detailed explanation of our probabilistic learning strategy, how to detect the overload on the system, and how to compute the amount of load to be dropped in order to meet the given latency bound. Finally, we explain the functionality of the load shedder.

\subsection{The \framework~Architecture}  
To enable load shedding, we extend the architecture of a CEP operator by adding the following components--overload detector, utility models, and load shedder (LS)--as depicted in Figure \ref{fig:ls}. 

The overload detector detects if there exists an overload on the operator. It checks the input event queue size periodically where the incoming windows of primitive events are queued. In case of an overload, the LS drops primitive events from windows to prevent the violation of the defined latency bound ($LB$). The utility models contain the utility of primitive events in a window and other information that is needed by the LS.   

Now, we explain how these 3 components are related. Upon detecting an overload, the overload detector commands the LS to drop events. On receiving this command, the LS uses the utility of primitive events in a window, available from the utility models, to decide on which events to drop.

Please note that load shedding is a time-critical task where it directly affects the CEP system performance and hence it must be lightweight and efficient. As we will see later, our load shedder has very low overhead. Contrarily, building the utility models can afford to be computationally heavy as it is not a time-critical task. 

\begin{figure}[t]
	\centering
	\resizebox{0.7\linewidth}{!}{

\newcommand{\gear}[6]{%
  (0:#2)
  \foreach \i [evaluate=\i as \n using {\i-1)*360/#1 }] in {1,...,#1}{%
   arc (\n:\n+#4:#2) {[rounded corners=1.5pt] -- (\n+#4+#5:#3)
    arc (\n+#4+#5:\n+360/#1-#5:#3)} --  (\n+360/#1:#2)
  }%
  (0,0) circle[radius=#6] 
}

\tikzset{label style/.style={font=\fontsize{24}{26}\selectfont\bf, color=black},
	rectangle style/.style ={color= blue, line width=4pt }, 
	arrow style/.style={color=black, line width=2pt, font=\fontsize{24}{26}\selectfont\bf, -{>[scale=2.5, length=7, width=4]}}, 
	instance style/.style={color=blue, line width=3pt} }

\begin{tikzpicture}[]

\draw  [arrow style, label style ] (2.5, 3) -- (5.5,3) node[pos=0.4, below, align= left] {windows};

\newcommand\XStart{5.5}
\newcommand\YStart{2.5}
\newcommand\SquareSize{1}
\foreach \n in {0,...,4}
	\draw[rectangle style, rounded corners] (\n + \XStart, \YStart) rectangle (\n + \XStart + \SquareSize, \YStart+ \SquareSize);
	
\node [label style ] at (8, 4) {input queue};

\draw  [arrow style, label style ] (10.5, 3) -- (12, 3);
	
\draw [instance style](14, 3 ) ellipse [x radius=3cm,y radius=2.4cm]; 
\begin{scope}[xshift=15.25cm, yshift=3cm]
	\fill[even odd rule]  \gear{8}{0.5}{0.75}{10}{2}{0.25};
\end{scope}
\node[label  style, below=2pt] at (15.15, 2.3) {process};
\node[label  style, above] at (15, -0.5) {operator};
\draw [rectangle style, red] (12, 3.75) -- (13.5, 3.25) -- (13.5, 2.75) -- (12, 2.25) --cycle;
\path (13.5, 2.75) -- (12, 2.25) coordinate[pos=0.5] (LSMidBelow);
\path (12, 3.75) -- (13.5, 3.25) coordinate[pos=0.5] (LSMidAbove);
\node[label  style] at (12.75, 3) {LS};
\draw  [arrow style, label style ] (13.5, 3) -- (14.5, 3);
\draw  [arrow style, label style ] (16, 3) -- (20, 3) node[pos=0.7, above=4pt, align= left] {complex\\ events};

\draw[rectangle style] (4.9,-1.5) rectangle (8.5, 0.5)  node [pos=.5,  label style, align=left]{overload\\detector};
\draw[arrow style, label style] (6.75, 2.5)-- (6.75, 0.5);
\draw  [arrow style, label style ] (8.5, -0.5) -- (12.75, -0.5) node[pos=0.6, align= left, below=2pt] {commands};
\draw[arrow style, label style] (12.75, -0.5) -- (LSMidBelow) ;

\draw[rectangle style] (11.25, 6.5) rectangle (14.25, 8.5)  node [pos=.5,  label style, align=left]{utility\\ models};

\draw[arrow style, label style] (12.75, 6.5) -- (LSMidAbove) ;

\end{tikzpicture}}
	\caption{The \framework~Architecture.}
	\label{fig:ls}
	\vspace{-0.6cm}
\end{figure}


\subsection{Utility Models and Their Applications}
In this section, we explain, in detail, the utility models and the way they can be used to drop events.

\textit{\textbf{Utility Prediction Function.}}
\label{sec:u-prediction-overview}
The utility of a primitive event in a window is defined by its impact on the quality of results. As mentioned earlier, we represent utility as the probability of the primitive event to be part of a complex event.
Clearly, dropping primitive events that have a high probability to be part of complex events degrades the quality of results. Hence, we avoid dropping these primitive events by assigning high utility values to them.    
Please recall that we identified the \textbf{type} and \textbf{position} of an event within a window to be an indicator of whether or not this event has a high probability to contribute to detect a complex event. This implies that the type and position of an event determine its utility.

As a result, to map type and position of events to a utility value, we introduce the utility prediction function $U(T, P)$ that predicts the utility of a primitive event of type $T$ in the position $P$ within a window. As we will see later, this prediction function can be simply implemented based on statistical data collected from the operator.

\textbf{\textit{{Utility Threshold and Occurrences.}}}
\label{sec:uth-overview}
Upon receiving the drop command to drop $x$ events from each window, the LS must find those $x$ events that have the lowest utility values in a window. One simple approach is to \textit{sort} the utility values using an efficient sort algorithm. For example, heap sort has a time-complexity of $O(ws.log_2x)$, where $ws$ is the number of events in a window \cite{Skiena2008}.
However, this approach requires that the entire window is available to the LS before sorting of the utilities and consequently shedding of events are performed. But waiting until the arrival of all events of a window might introduce a high latency on event processing or might even cause violation of $LB$. Moreover, sorting needs to be performed in every window which might add additional overhead on the system that already suffers from an overload.

A better approach to avoid the above induced latency and overhead is to find a utility threshold  (denoted by $u_{th}$) that can be used on-the-fly to drop the desired number of events in a given window. In particular, we need a function that maps the number of events to drop per window ($x$) to a utility threshold $u_{th}$, i.e., f($x) \to u_{th}$.
To find the utility threshold $u_{th}$, we could predict the number of $x$ event occurrences in a window, whose utility is less or equal to the utility threshold $u_{th}$.


More specifically, in a window $w$, we define the number of event occurrences, whose utility is less or equal to a certain utility value $u$ as follows: $O(u)=|\{e: U(T, P) \le  u \}|$, where T is the type of event $e$ and P is its position in w.
The number of event occurrences $O(u)$ in a window $w$, as defined above, implicitly represents the cumulative occurrences of those utilities in $w$, whose values are less or equal to the utility value $u$ and hence, as a short hand, we call $O(u)$ as cumulative utility occurrences.

The utility threshold $u_{th}$ can be calculated using the inverse function of the cumulative utility occurrences $O(u_{th})$, where, given the number of events that should be dropped from each window, we can get the required utility threshold.

\textbf{\textit{{Applying Utility Models in Load Shedding.}}}
Now, we describe how load shedding is performed in \framework. To drop $x$ events from each incoming window, the LS first searches for the cumulative utility occurrences $O(u)$, which has a value $O(u) \ge x$. Then, the LS uses the utility value $u$ as a utility threshold $u_{th}$ to drop those $x$ events from each window.

To use the utility threshold $u_{th}$ and drop events, first, the LS  gets the next event $e$ from the input event queue of the  operator. Then, for each window $w$ to which the event $e$ belongs, the LS computes the utility value $u$ of the event $e$ in $w$ using the utility prediction function $U (T, P)$. If the event utility $u$ in the window $w$ is greater than the utility threshold $u_{th}$, the LS keeps the event $e$ in $w$. Otherwise,  it drops the event $e$ from $w$. The utility threshold $u_{th}$ enables the LS to drop $x$ events from each window.

\subsection{Model Building}
Having discussed the role of the utility prediction and the threshold prediction functions, in this section, we discuss, in detail, the manner in which we implement these functions.
For a clear explanation, let us introduce the following simple running example. We use a pattern matching query that considers a window of 5 events (i.e., window size = 5) and an input event stream consisting of only \textit{two} event types $A$ and $B$ (cf. Figure \ref{fig:running-example}).

\textbf{\textit{{Building the Utility Prediction Function.}}}
As mentioned in Section \ref{sec:u-prediction-overview}, the 
utility of a primitive event $U(T,P)$ is represented by the probability of the event to be part of the detected complex events. To predict the utility of primitive events in a window, we collect statistics, from the already detected complex events, on the types and relative positions within windows of \textit{primitive events} that contributed to those detected complex events. More specifically, we count the number of occurrences of each event type in each position in a window that contributed to the detected complex events. 
The number of occurrences of primitive events within detected complex events provides an incite into the importance (or utility) of the primitive event types and their relative positions within a window. As a result, we simply normalize those number of occurrences to generate the utility (i.e., $U(T, P)$) of a certain event type $T$ in a certain window position $P$.

These utility values are stored in a table called utility table (denoted by  $UT$). The utility table has $(M x N)$ dimensions, where $M$ represents the number of different event types and $N$ represents the window size $ws$. Each of its cells $UT(T, P)$ represents the utility of a specific \textit{event type} $T$ in a certain \textit{position} $P$ in a window, where the utility value $U(T, P)$ of an event is stored in $UT(T, P)$.

The values in $UT$ could be too fine-grained which can cause more memory and computational overhead. To avoid this overhead, we limit the number of different utility values by multiplying each cell value in $UT$ by 100 and rounding it to an integer, i.e., $UT(T, P) \in [0, 100]$.
Referring to our above example, Table \ref{tab:ut} shows a utility table that is generated from the collected statistical data. 

\renewcommand{\arraystretch}{0.6}
\begin{figure}
	\begin{minipage}[t]{.48\textwidth}
		\centering
		\begin{tabular}{ | c| c | c |c |c |c |}
			\hline
			\diagbox[width=7em, innerleftsep=2pt]{event type}{position} & 1 & 2 & 3 & 4 & 5 \\ \hline
			$A$ & 70 & 15 & 10 & 5 & 0 \\ \hline
			$B$ & 0 & 60 & 30 & 10 & 0 \\ 
			\hline
		\end{tabular}
		\captionof{table}{$UT$ generated from the collected statistical data. }
		\label{tab:ut}
	\end{minipage} 
	\hfill
	\begin{minipage}[t]{.48\textwidth}
		\centering
		\resizebox{0.60\linewidth}{!}{
\pgfplotsset{every axis/.append style={
		axis x line=left,    
		axis y line=left,    
		axis line style={draw=none},
		tick style={draw=none},
		xlabel={$x$},          
		ylabel={$y$},          
		label style={font=\fontsize{24}{26}\selectfont\bf , color=black},
		x label style={ below=4mm},
		y label style={ above=3mm},
		tick label style={font=\fontsize{21}{24}\selectfont\bf, color= black},
		every node near coord/.append style={font=\fontsize{24}{21}\selectfont\bf, color= black},
	}}

\begin{tikzpicture}[]
\begin{axis}
	[
	ybar,
	 x=1.2cm, 
	bar width=.7cm,
	color=black,
	fill=black,
	xlabel=	cumulative utility occurrences $O(u)$,
	ylabel= utility threshold $u_{th}$,
	xmin=0.5,
	xmax=7.5,
	ymin=0,
	ymax=80,
	ymajorgrids,
	xtick=data,
	xticklabels={{1.2},{1.4},{2.3},{2.8},{3.7},{4.2},{5}},
	bar shift=0pt,
	nodes near coords,
	]
	\addplot [draw=black,fill=blue]  table [x index=0, y index=2, col sep=comma] {./figures/tikz/cdt-correct.csv};
\end{axis}

\end{tikzpicture}}
		\vspace{-0.4cm}
		\captionof{figure}{$CDT$ computed from Table \ref{tab:ut} ($UT$) and the predicted position shares in a widnow. }
		\label{fig:cdt-correct}
	\end{minipage}
	\vspace{-0.4cm}
	\caption{Simple running example.}
	\label{fig:running-example}
	\vspace{-0.6cm}
\end{figure}

\renewcommand{\arraystretch}{1}

\textbf{\textit{{Building Utility Threshold and Occurrences.}}}
As we discussed in Section \ref{sec:uth-overview}, to drop $x$ events from each window, we should find a utility threshold $u_{th}$ that results in dropping $x$ events from each window, where the utility threshold $u_{th}$ is the inverse function of the cumulative utility occurrences $O(u_{th})$. 
In particular, we should find a utility value $u$ that is greater or equal to the utility value of $x$ events in a window, i.e., $O(u) \ge x$. Then, we use $u$ as a utility threshold $u_{th}$ to drop $x$ events from each window.

To find the utility threshold $u_{th}$, we need to calculate the cumulative utility occurrences $O(u)$ in a window. Since the utilities of events of all types and in all positions in a window are stored in $UT$, we can determine the cumulative utility occurrences $O(u)$ from $UT$. The cumulative utility occurrences depend on the distribution of utilities within windows captured in $UT$.

To predict the utility threshold $u_{th}$, let us, for now, assume that there is only \textbf{one event type} in the input event stream (i.e., $M= 1$) and hence the dimensions of $UT$ become ($1xN$), recall $N$ is the number of positions (i.e., events) in a window.
Since the utility values in $UT$ are between 0 and 100 (recall that $UT(T, P) \in [0, 100]$), there will be a maximum of 101 different utility values, where each utility value in $UT$ may repeat several times. To build the cumulative utility occurrences $O(u)$ for each individual utility value $u \in [0, 100]$, we first count the number of occurrences $o_u$ of each individual utility value $u \in UT$. 
Once we have the occurrences of each utility value, we can calculate the cumulative utility occurrences.
To this end, the number of occurrences $o_u$ of the individual utility values $u$ are accumulated together in a cumulative distribution fashion as follows:
\begin{equation}
\begin{aligned}
 O(u)= 
	 \begin{cases}
		 o_u,&  \text{if } u=0 \\
		 o_u + O(u-1), & \text{otherwise}
	 \end{cases}
\end{aligned}
\label{eq:occurrences}
\end{equation}

So far, we have assumed that there exists a \textbf{single event type} in the input event stream. However, there may be \textbf{multiple event types} in the input event stream. In this case, for \textit{each single position} in $UT$, there exists a utility value for each event type. 
For example, in Table \ref{tab:ut}, each single position in $UT$ has two utility values, one utility value for the event type $A$ and one for the event type $B$. In the table, $UT(A, 1)= 70$ and $UT(B, 1)= 0$. 
This means that a \textit{single position} in $UT$ is incrementing the occurrences of  multiple utilities. As a result, to count the number of utility occurrences $o_u$, we need to consider each position in $UT$ as a shared position between all event types. More specifically, for each event type, we count a utility occurrence $o_u$ in a certain position in $UT$ as a \textit{fractional value} instead of counting it as a \textit{full occurrence}. We call these fractional values as \textit{position shares in a window}. 
We could predict the position shares in a window between different event types from the distribution probability of the primitive events within the window. The position shares in a window $S(T, P)$ of a primitive event of type $T$ at position $P$ in the window equals the probability of this event type $T$ to come at the position $P$ in the window.

Now, to compute the cumulative utility occurrences $O(u)$ in case of \textbf{many event types},  we count the occurrences $o_u$ of the utility value $u$ in $UT$ as a fractional value by its corresponding values from the position shares in a window. For each utility value $u= UT(T, P)$ for the event type $T$ at position $P$ in $UT$, we increase the number of occurrences $o_u$ by $S(T, P)$. The cumulative utility occurrences $O(u)$ is then computed as in case of a single event type using equation \ref{eq:occurrences}.

We store the cumulative utility occurrences $O(u)$ in an array called $CDT$, where the utility values $u$ are used as indices and the cumulative utility occurrences $O(u)$ are used as the actual values, i.e., $CDT(u)= O(u)$.
$CDT$ is a single dimensional array of size (101), which is the maximum number of different utility values in $UT$. An \textit{index} $u$ in $CDT$ represents a utility value $u$ in $UT$ and its cell value $CDT(u)$ represents the cumulative utility occurrences $O(u)$ of the utility value $u$. 

Since the utility threshold $u_{th}$ is the inverse function of the cumulative utility occurrences $O(u)$, we extract $u_{th}$ from $CDT$.
To find a utility threshold $u_{th}$ which drops $x$ events from each window, we iterate over $CDT$ to find a cell value $CDT(u)$ that is $\ge x$, which means that the number of primitive events with utility values less or equal to $u$ occur at least $x$ times in each window. Hence, using $u$ as a utility threshold drops at least $x$ events from each window. 
We explain utility threshold prediction further with the help of our running example.
Figure \ref{fig:cdt-correct} shows the $CDT$ computed from $UT$ in Table \ref{tab:ut} and the predicted position shares in a window. 
Now, to drop $x= 2$ events from each window, in the figure, $CDT(10)= 2.3 > x$. Thus, to drop $x=2$ events from each window, we use the utility threshold $u_{th}= 10$.


Algorithm \ref{alg:cdt} explains the construction of $CDT$ from both $UT$ and the predicted position shares in a window. The algorithm first counts the number of occurrences $o_u$ of each individual utility value $u$ in $UT$ (cf. lines (2-5)).  
It iterates over each cell in $UT$ (cf. lines (2-3)) and gets its value $u= UT(T, P)$, i.e., the utility of the event type $T$ at the position $P$ in the window (cf. line 4). Then, in line 5, the algorithm increments the cell value in a temporary array $temp$ which is at index $u$ by the fractional value $S(T, P)$.  
Since the utility values are used as indices in $CDT$, they are already sorted in an ascending order. Finally, the algorithm accumulates the values in $CDT$ starting from index 0 where $CDT(u)= CDT(u) + CDT(u-1),\ u= 1..100$ (cf. lines(7-9)). 
\begin{algorithm}
	\setbox0\vbox{\small
		{\fontsize{8.0}{9.0}\selectfont
			\begin{algorithmic}[1]
				\algsetblockdefx[function]{func}{endfunc}{}{0.2cm}[3]{#1 \textbf{#2} (#3) \textbf{begin}}{\textbf{end function}}
				
				\func {} {computeCDT} {}
				
				\For {$T=0\quad to\quad (M -1)$} \Comment {$M$: Number of different event types}
				\For {$P=0\quad to\quad (N -1)$} \Comment {$N$: the window size $ws$}				
				\State $\mathit{u} =  \mathit{UT(T, P)}$
				\State $\mathit{temp(u)} \mathrel{+}=  S(T, P)$ \Comment {$temp(u)= o_u$} 
				\EndFor
				\EndFor	
				
				\State \textit{//accumulate utility values in an ascending order.}
				\State $CDT(0) = temp(0)$ 
				\For {$u= 1\quad to\quad 100$}
				\State $CDT(u) = temp(u) + CDT(u -1)$ 
				\EndFor	
				
				\endfunc
			\end{algorithmic}
		}
	}
	\centerline{\fbox{\box0}}
	\caption{Building CDT table.}
	\label{alg:cdt}
\end{algorithm}
\vspace{-0.3cm}

Please note that building $UT$ and $CDT$  is a continuous task  where they are periodically updated .


\subsection{\textbf{Overload Detection}}
Having explained the way utility models are built, we now provide details on when the LS should drop events and how many and in which interval should events be dropped.

To detect an overload on an operator, the overload detector periodically monitors the input event queue and  calculates the estimated latency for the incoming events (denoted by  $l(e)$). It compares $l(e)$ with the defined latency bound $LB$ and decides to drop events if $LB$ might be violated.

The value of $l(e)$ depends on event processing latency (denoted by $l(p)$) and event queuing latency (denoted by  $l(q)$), in fact,  
$l(e) = l(q) + l(p)$. 
Event processing latency $l(p)$ represents the time an event needs to be processed by the operator in all windows.  $l(p)$ is calculated from the throughput  of the operator (denoted by $th$). The throughput $th$ represents the maximum number of events the operator can process per second. Event queuing latency $l(q)$ represents the time an event must wait before it gets processed by the operator. This time depends on the number of queued events $n$ before this event $e$ in the input event queue and on $l(p)$, i.e., $l(q) = n * l(p)$.
This means that an event $e$ in position $n$ in the input event queue has an estimated latency $l(e)= (n - 1) * l(p) + l(p) = n * l(p)$.  

From the given latency bound $LB$ and the event processing latency $l(p)$, we can get the maximum allowed queue size (denoted by $q_{max}$) before violating $LB$, where $q_{max} = {LB}/{l(p)}$.

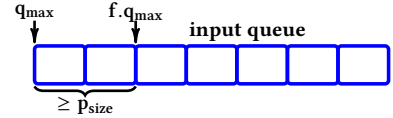
\begin{figure}[t]
	\centering
	\resizebox{0.6\linewidth}{!}{

\tikzset{label style/.style={font=\fontsize{24}{26}\selectfont\bf, color=black},
	rectangle style/.style ={color= blue, line width=4pt }, 
	arrow style/.style={color=black, line width=3pt, font=\fontsize{24}{26}\selectfont\bf, -{>[scale=2.5, length=7, width=4]}}, 
}

\begin{tikzpicture}[]

\newcommand\XStart{0}
\newcommand\YStart{0}
\newcommand\xSquareSize{2}
\newcommand\ySquareSize{1.5}
\newcommand\numCells{6}
\foreach \n in {0,...,\numCells}
\draw[rectangle style, rounded corners] (\xSquareSize*\n + \XStart, \YStart) rectangle (\xSquareSize*\n + \XStart + \xSquareSize, \YStart+ \ySquareSize);
	
\node [label style, above right, ] at (\xSquareSize*\numCells/2, \ySquareSize+ 0.1) {input queue};

\draw  [arrow style, label style ] (\XStart, \ySquareSize + 1) -- (\XStart, \ySquareSize+0.1) node[ pos= -.40] {$\mathbf{q_{max}}$};
\draw  [arrow style, label style ] (\XStart + \xSquareSize*2 ,\ySquareSize + 1) -- (\XStart + \xSquareSize*2 , \ySquareSize+0.1) node[ pos= -.40] {$\mathbf{f.q_{max}}$};

 \draw[label style, line width=3pt,decorate,decoration={brace,amplitude=10pt, mirror}] (\XStart,-0.1) -- (\XStart + \xSquareSize*2,-0.1) node[midway, below=8pt] {$\mathbf{\ge p_{size}}$};

\end{tikzpicture}}
	\vspace{-0.2cm}
	\caption{Partition Size.}
	\label{fig:partition-size}
	\vspace{-0.4cm}
\end{figure}

Waiting until the queue size (denoted by  $q_{size}$) equals $q_{max}$ to start dropping events might be too late and can cause $LB$ violation. Therefore, we start dropping events, if the following inequality holds: $q_{size} > f.q_{max}$, where $f \in [0,1]$,  see Figure \ref{fig:partition-size}. 
A high $f$ value, on one hand, avoids unnecessarily dropping events-- in cases the events are only queued for a short time as in short burst situations. But on the other hand, it might force the LS to drop events with high utility values to avoid $LB$ violation-- in case the queue size gets close to $q_{max}$. Later, we explain how to choose an appropriate $f$ value.

\textbf{\textit{Dropping Interval.}}
So far, we have considered dropping $x$ events per window. 
However, the window size might not be the best dropping interval to meet the given latency bound $LB$.
The reason is that as the LS starts dropping events when $q_{size} > f. q_{max}$, the buffer that we have before violating the latency bound ($LB$) is of size ($q_{max} - f.q_{max}$) events (cf. Figure \ref{fig:partition-size}). More specifically, we need to drop $x$ events from at least  every ($q_{max} - f.q_{max}$) events (i.e., dropping interval) in order to meet LB. Therefore, please note that the dropping interval must be less or equal to ($q_{max} - f.q_{max}$).

As a result, if the window size $ws$ is less or equal to this buffer size (i.e., $q_{max} - f.q_{max}$), then the interval of dropping $x$ events is preserved and the utility threshold $u_{th}$ can be calculated for the entire window. However, if the window size $ws$ is greater than the buffer size, there is a risk of $LB$ violation, especially if the utility values are not evenly distributed in windows, e.g., all events with high utilities come together in a certain region of the window. In this case, the utility threshold $u_{th}$ will result in dropping $x$ events from each window but not necessarily from each dropping interval (i.e., the buffer size) if the size of the high utility region of the windows is greater than the buffer size. This might result in $LB$ violation.


Therefore, we must partition the window into smaller partitions of size less or equal to the buffer size, i.e., $q_{max} - f.q_{max}$ (as can be seen in Figure \ref{fig:partition-size}) and  drop $x$ events from each partition.
While the partition size cannot be greater than the buffer size (cf. the above mentioned constraint), of course, it can be less than the buffer size. However, larger the partition size is, greater is the probability to find low utility values to drop, resulting in better quality. As a result, we try to use a partition size which is as large as possible (of course, the upper bound being the buffer size). 
More specifically, we partition a window in $\rho$ partitions of equal sizes, where $\rho= ceil(\frac{ws}{q_{max} - f.q_{max}})$. As a result, the partition size $p_{size}= \frac{ws}{\rho}$. We use the partition size as a \textit{dropping interval} in which $x$ events should be dropped. Therefore, we cannot use the utility threshold $u_{th}$ that comes from a full window but instead we have to use a utility threshold $u_{th}$ for each partition in order to drop $x$ events in each dropping interval.

We already discussed how to compute CDT, i.e., the cumulative utility occurrences $O(u)$, for a complete window. However, since a window might be divided into more than one partition (when $\rho > 1$), we must compute for each partition its own $CDT$. Please note that $UT$ will be calculated as before. However, the utility threshold $u_{th}$ needs to be calculated based on the partition size $p_{size}$ within which shedding must be performed. 
Therefore, we compute $CDT$  for each partition of size $p_{size}$ within $UT$. So, now, to drop $x$ events from each partition of the incoming windows, each partition has its own utility threshold $u_{th}$. 


\textbf{\textit{Dropping Amount.}}
The dropping amount represents the number of primitive events that must be dropped from each partition of each window.
Determining how many events $x$ to drop per partition depends on the input event rate and the operator throughput $th$. The overload detector computes the difference $\delta$ between the input event rate $R$ and $th$, where $\delta= R - th$, to get the extra incoming events per \textit{second}. Then, the number of events $x$ to drop per partition is computed as follows: $x = \delta . \frac{p_{size}}{R}$, where $\frac{p_{size}}{R}$ represents the partition size in seconds. 

\textbf{\textit{Appropriate f Value.}}
As we mentioned above, using a high $f$ value prevents dropping events in short burst cases and hence decreases the degradation in the quality of results. However, the $f$ value controls the partition size $p_{size}$, and hence using a high $f$ value forces us to use a small partition size to avoid $LB$ violation. A small partition size might result in dropping events that have high utility values. That can happen if all events in a partition have high utility values. Therefore, we should choose a minimum $f$ value that still allows to have a partition size which avoids dropping high utility events. 

Fortunately, we already have the distribution of utilities within a window captured in $UT$. We can take advantage of this knowledge to determine $f$ value.  To find the $f$ value, we propose to cluster the utilities in $UT$ into several classes of importance. The goal is to partition the windows depending on the $f$ value into one or more partitions, where in each  partition, there exists at least $x$ events from the low utility classes. This way, in each partition, the low utility events can be dropped, hence reducing the degradation in the quality of results. Therefore, we can choose the $f$ value that ensures the above partition size.

\subsection{\textbf{Load shedding}}
Now, we explain, in detail, the functionality of \framework's load shedder (LS) component.
Events are dropped from individual windows without affecting other windows. An event might be dropped from one window while it is still there in other windows as the event utility $U(T, P)$ can be different in different windows, since the event  position $P$ is different in different windows.
The LS checks for each incoming event in a window and decides on whether or not to drop it depending on the event utility in $UT$ and on the utility threshold $u_{th}$ of the partition to which the event belongs. Hence, the LS must be lightweight since it is performed for every event in a window.

Upon receiving the drop command from the overload detector, the LS searches for the utility thresholds corresponding to each partition of an incoming window. Note that the entire window can also be a single partition, i.e., there is only one partition ($\rho =1$), if $ws \le (q_{max} - f.q_{max})$ (cf. previous section). Having noted the utility thresholds for every partition of a window, the LS proceeds to drop events from the incoming windows.
So, for each event $e$ in a window, the LS gets its utility $U(T, P)$ from $UT$ and also determines the partition ($part$) in a window to which the event $e$ belongs, both in O(1) time-complexity. Then, the LS compares the event utility $U(T, P)$ with the utility threshold $u_{th}$ of the corresponding partition ($part$) to decide on whether or not to drop $e$ from the window. If the utility of the event is less or equal to the utility threshold of its corresponding partition, the LS drops the event.

Algorithm \ref{alg:loadShedder} explains the LS functionality more formally. If $q_{size} > f. q_{max}$, the overload detector requires the LS to activate the shedding. It also sends drop commands which contain the number of events $x$ to drop per partition to LS.  
The LS receives drop commands from the overload detector where it firstly calculates the utility threshold $u_{th}$ for each partition depending on the required number of events $x$ to drop per partition (cf. lines 1-7).  To calculate the utility threshold $u_{th}(part)$ for each partition $part$, the LS iterates (cf. lines 2-3) over its corresponding $CDT$ to search for a value $CDT(part, u)$ which is  $\ge x$ (cf. line 4). Then, the index $u$ of this value $CDT(part, u)$ is used as the utility threshold $u_{th}(part)$ for this partition $part$ (cf. line 5).


In case load shedding is active, for each event $e$ in the incoming windows, the LS checks if it needs to drop the event $e$ (cf. lines 8-17). First, the LS finds the partition in the window to which the event $e$ belongs (cf. line 12).
Then, the LS checks if the utility value $UT(T, P)$ of this event $e$ in $UT$ is less or equal to the utility threshold $u_{th}(part)$ of its calculated partition (cf. lines 13-16)---just a simple lookup in $UT$. It then returns true if the event should be dropped, otherwise false. This shows that our load shedder is extremely lightweight and it takes the shedding decision in O (1) time-complexity for each event in a window.


\begin{algorithm}
	\setbox0\vbox{\small
		{\fontsize{8.0}{9.0}\selectfont
			\begin{algorithmic}[1]
				\algsetblockdefx[function]{func}{endfunc}{}{0.2cm}[3]{#1 \textbf{#2} (#3) \textbf{begin}}{\textbf{end function}}
						
				\func {}{getUtilityThresholdForEachPartition}{$x$} 
				\For {$\mathit{part}=0\quad to\quad (\rho -1)$}
					\For {$u=0\quad  to\quad 100 $}
						\If {$\mathit{CDT(part, u)} \ge x$}
							\State $\mathit{u_{th}(part)} =  u$
							\State $\mathbf{break}$ {\fontsize{7.0}{8.0}\selectfont \Comment {break the inner  loop and proceed to the next partition.}}
						\EndIf
					\EndFor
				\EndFor						
				\endfunc
				\hfill
				\func {}{applyLS}{$T, P$} {\fontsize{7.0}{8.0}\selectfont \Comment $T$: event type \& $P$: event position in the window}			
				\If{$!\mathit{LS.active}$}
					\State $\mathbf{return} \quad false$
				\Else
					\State $\mathit{part} =  \frac{n}{\mathit{p_{size}}}$
					\If {$\mathit{UT(T, P)} \le \mathit{u_{th}(part)}$}
						\State $\mathbf{return} \quad true$
					\Else
						\State $\mathbf{return} \quad false$	
					\EndIf	
				\EndIf		
		
				\endfunc
				
			\end{algorithmic}
		}
	}
	\centerline{\fbox{\box0}}
	\caption{Load shedder.}
	\label{alg:loadShedder}
\end{algorithm}
\vspace{-0.3cm}

\subsection{Extensions}
In this section, we explain three extensions to our load shedding approach---handling variable window size, using bins for large windows, and retraining the model. Handling variable window size enables our approach to work with windows of different sizes. While, the use of bins optimizes our approach and enables it to work with large windows.

\textbf{\textit{Handling Variable Window Size.}}
\label{sec: variable-windows}
The incoming windows might have a variable window size $ws$ depending on the window splitting strategies. As mentioned above, in CEP systems, there exist three main window splitting strategies---count-based, time-based and pattern-based. In count-based, $ws$ is always fixed while in time-based and pattern-based, $ws$ might change depending on the input event rate or content of the events \cite{Mayer:2017:MCO:3093742.3093914}.

As explained earlier, in order to implement the utility prediction function, we use $UT$ which has a fixed number of event positions $N$, where $N= ws$. However, if the window size $ws$ varies and is not fixed, we need a way to find $N$. Therefore, to handle variable window size, we profile the operator and choose $N$ as the average seen window size. Since $N$ might be different from the incoming window sizes, in the following, we explain the required modifications to our approach during both model building and load shedding to incorporate variable window size.

\paragraph{During Model Building:}
We need a way to map the event positions in windows to the event positions in $UT$ that has a fixed number of positions $N$. 
To do that, we normalize the size of incoming windows to $N$. For each incoming window, if $ws > N$, we scale down $w$ where more than one position in $w$ is mapped to a single position in $UT$. On the other hand, If $ws < N$, we scale up $w$ where each position in $w$ is mapped to one or more positions in  $UT$. The scaling factor $sf$ can easily be computed as follows: $sf= \frac{ws}{N}$. For example, let $N=100$ and $ws=200$, then $sf=\frac{200}{100} = 2$. This means that every two positions in $w$ is mapped to a single position in $UT$. 

\paragraph{During Load Shedding:}
The window size may also vary while performing  load shedding. So, in this case while processing every incoming event $e$ of a window $w$, the LS must determine the relative position of the event $e$ within the window $w$, instead of the exact position. In this way, the LS can map the learned utility values in $UT$ to the event $e$. To map relative position of the event $e$ in the window $w$ to the exact position in $UT$, we again scale down $ws$ if $ws > N$ and scale up  $ws$ if $ws < N$. Since during scaling up $ws$, an event $e$ in $w$ is mapped to more than one cell in $UT$, the utility of $e$ is the average value of all these cell values in $UT$. 

As mentioned above, the cumulative utility occurrences $O(u)$, which are stored in $CDT$, are computed from $UT$ that has a fixed number of positions $N$. In case of varying window sizes, the utility threshold $u_{th}$ is calculated from $CDT$ without any modification. This is because, the utility values in $UT$ already capture the variation in the window size. So, the calculated utility threshold $u_{th}$ from $CDT$ implicitly scales up/down depending on the window size.

The problem  with variable window size during load shedding is that we process events on their arrival without waiting until the end of the windows. Thus, the incoming window size is not known at the time when the LS performs lookup in $UT$ to get the utility of an event in a window based on its relative position. It is not possible to get a relative position if the incoming window size is not known.  However, the window size is important for the lookup and we must predict it. For example, in case of time-based window, the event input rate could be used to predict the window size. Predicting the window size is already researched in literature \cite{Mayer:2017:MCO:3093742.3093914} and will not be the focus of this paper.    

\textbf{\textit{Using Bins for a Large Window Size.}}
The average seen window size $N$ might be too large. This might result in a huge size of $UT$, thus wasting computing resources. Therefore, bins of size $bs$ are used to map several neighboring positions for each specific event type in a window to one single position in $UT$, thus reducing its size. In Section \ref{sec:results}, we discuss more about the impact of the bin size on the quality of results.

\textbf{\textit{Model Retraining.}}
If the distribution of the input event stream changes over time, the constructed model becomes inaccurate, which may adversely impact the quality of results. Therefore, in this case, we must retrain the model to capture the changes in the input event stream. We can either periodically retrain the model to capture these changes  or we can use a statistical approach that triggers the need to retrain the model (we leave this approach for future work). 

\section{Performance Evaluations}
\label{sec:results}
Next, we evaluate the performance of \framework~by analyzing its impact on the quality of results when the input event rate exceeds the operator throughput $th$.
\subsection{Experimental Setup}
Here, we describe the evaluation platform, the baseline implementation, datasets and queries used in the evaluations.

\textbf{\textit{Evaluation Platform.}}
We run our evaluation on a machine which is equipped with 8 CPU cores (Intel 1.6 GHz) and a main memory of 24 GB. The  OS used is CentOS 6.4. We run the operator in a single thread which is used as a resource limitation. Please note, that the performance of \framework~is independent of the parallelism degree of the operator.
We implemented \framework~by extending a prototype CEP framework which is implemented using Java.

\textbf{\textit{Baseline.}}
As evaluation results showed that a completely random event shedder is comprehensively outperformed by \framework, we did not think it would be a fair comparison and therefore looked at comparing \framework~with a baseline that is similar to state-of-the-art solutions.
We implemented a load shedding strategy which is similar to the strategy in \cite{3:He2014OnLS} as a baseline strategy (denoted by BL). Our implemented baseline strategy also captures the notion of weighted sampling techniques in stream processing \cite{3:Tatbul:2003:LSD:1315451.1315479}.  BL assigns utility values to the primitive events in a window depending on the repetition of those primitive events in the pattern and on their frequencies in windows. An event type receives a higher utility proportional to its repetition in a pattern. Depending on the event utility, BL decides the amount of events that should be dropped from each event type in a window, where it uses uniform sampling to drop those required amounts from each event type. BL, as in \cite{3:He2014OnLS}, does not consider the order of events in a pattern and in input event streams. 

\textbf{\textit{Datasets.}}
We employ two different datasets from real-world.
First, we use a stock quote stream from the New York Stock Exchange (NYSE). 
This dataset contains real intra-day quotes of 500 different stocks from NYSE collected over two months from Google Finance \cite{google_finance}. The quotes have a resolution of 1 quote per minute for each stock symbol. We refer to this dataset as the \emph{NYSE Stock Quotes} dataset. 
Second, we use a position data stream from a real-time locating system (RTLS) in a soccer game \cite{Mutschler:2013:DGC:2488222.2488283}. Players, balls and referees (called objects) are equipped with sensors that generate events which contain their position, velocity, etc. The sensor data are generated at a high rate causing high redundancy. Thus, we filter redundant events and keep only one event per second for each object. We refer to this dataset as the \emph{RTLS} dataset.

\textbf{\textit{Queries.}}
We employ four queries (Q1, Q2, Q3, Q4) that cover an important set of operators in CEP: sequence operator, sequence with any operator, and sequence operator with repetition, all with skip-till-next/any-match \cite{Zhang:2014:COE:2588555.2593671, 1:ch1994snoop, 1:Balkesen:2013:RRI:2488222.2488257, 1:cu2010tesla, Dayarathna:2018:RAE:3186333.3170432, Wu:2006:HCE:1142473.1142520}. Moreover, the queries use both \textbf{time-based} and \textbf{count-based} sliding window strategies with \textbf{different predicates}. Furthermore, to study the robustness of our load shedding with different \textit{selection policies}, we 
implemented all queries with the \textit{first} and \textit{last} selection policies. 
All queries skip the intermediate not matching primitive events, i.e.,  skip-till-next/any-match. 
The queries are as follows :

Q1 (sequence with any operator): uses the RTLS dataset. It detects a complex event when any $n$ defenders of a team (defined as $\mathtt{DF}$) defend against a striker (defined as $\mathtt{STR}$) from the other team within $\mathit{ws}$ seconds from the ball possessing event by the striker. The defending action is defined by a certain distance between the striker and the defenders. We use two players as strikers; one striker from each team. Q1 is of form: seq ($STR$; any ($n$, $DF_1$, $DF_2$, .., $DF_n$)), where $DF_x$ is the defend event of the player $x$.

Q2 (sequence with any operator): is adopted from related work \cite{spectre:2017}. It detects a complex event when any $n$ rising or any $n$ falling stock quotes of any stock symbol (defined as $RE$ or $FE$, respectively) are detected within $\mathit{ws}$ seconds from a rising or falling quote of a leading stock symbol (defined as $MLE$). The leading stock symbols are composed of a list of 5 technology blue chip companies. Q2 is of form: seq ($MLE$; any ($n$, $RE_1$, $RE_2$, ..,$RE_n$)) or seq ($MLE$; any ($n$, $FE_1$, $FE_2$,..,$FE_n$)), where $RE_x$ or $FE_x$ is  rising or falling event of the stock company $x$.

Q3 (sequence operator): detects a complex event  when rising or falling stock quotes of 20 certain stock symbols (defined as $RE$ or $FE$, respectively) are detected within $\mathit{ws}$ events in a certain sequence.  Q3 is of form: seq ($RE_1$; $RE_2$;..;$RE_{20}$) or seq ($FE_1$; $FE_2$;..;$FE_{20}$), where $RE_x$ and $FE_x$  are defined as in Q2. 

Q4 (sequence operator with repetition): detects a complex event when 10 rising or 10 falling stock quotes of certain stock symbols (defined as $RE$ or $FE$, respectively) \textit{with repetition} are detected within $\mathit{ws}$ events in a certain sequence. Q4 is of form: seq ($RE_1$;  $RE_1$; $RE_2$; $RE_3$; $RE_2$; $RE_4$; $RE_2$; $RE_5$; $RE_6$; $RE_7$; $RE_2$; $RE_8$; $RE_9$; $RE_{10}$), where $RE_x$ is defined as in Q2. The sequence for falling quotes is similar. 

\subsection{Experimental Results}
In this section, first, we evaluate the impact of our probabilistic LS strategy (\framework) on the quality of results, particularly the number of false positives and false negatives, and compare its results with the results of BL.  Then, we show the impact of variable window size and bin  size on the quality of results and analyze the overhead of the LS.

If not noted otherwise, we employ the following settings. The number of complex events per window is one. For Q1 and Q2, we use a \textit{time-based} sliding window. In both queries, a \textit{logical} predicate is used to open new windows. For Q1, a new window is opened for each incoming striker event (STR). While for Q2, a new window is opened for each incoming event of the leading stock symbols (MLE). Q3 and Q4 use \textit{count-based} sliding window. In Q3, we also use a \textit{logical} predicate where a new window is opened for each incoming event of the leading stock symbols (MLE). For Q4, a \textit{count-based} predicate is used where a new window is opened every 100 events, i.e., the slide size equals 100 events.
We use a latency bound $LB = 1$ second and an $f$ value $= 0.8$. Moreover, we stream the datasets from stored files to the system with an event input rate that is less or equal to the maximum processing rate of the operator (i.e., the operator throughput $(th)$) until the model is built. After that, we increase the input event rate to a rate that is higher than the operator throughput $th$ by 20\% (denoted by $R1$) and 40\% (denoted by $R2$).
We execute several runs for each experiment and show the mean value and standard deviation.

\textbf{\textit{Impact on the quality of results and the given latency bound.}}
\label{sec:quality-fpn}
First, we show the impact of \framework~on the given latency bound and on the quality of results, i.e., the number of false negatives and false positives. The quality is influenced by the input event rate and the ratio of pattern size to window size. Therefore, we show results using different input event rates and different ratios of pattern size to window size.

To evaluate the performance of \framework, we run experiments with Q1, Q2, Q3 and Q4. For Q1 and Q2, we use a variable pattern size with a fixed window size to get different ratios of pattern size to window size.
For Q1, we use a window size $ws$ of  15 seconds ($\approx$ 700 events) and the following pattern sizes (i.e., number of defenders):  $n$= 2, 3, 4, 5, 6. The window size $ws$ for Q2  is 240 seconds ($\approx$ 2000 events) and the pattern is of the following sizes: $n$= 10, 20, 30, 40, 50, 60, 70, 80. Since the pattern size is fixed in Q3 and Q4, we use a variable window size to get different ratios of pattern size to window size.  For both Q3 and Q4, we use the following window sizes: $ws= $300, 600, 1200, 1500, 1800, 2000 events.

\begin{figure*}[t]
	\centering
	\begin{subfigure}[t]{0.70\linewidth}
		\centering
		\includegraphics[width=0.75\linewidth]{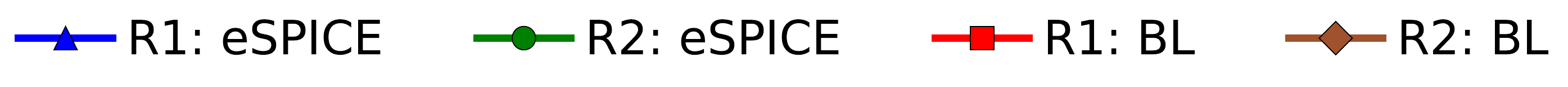}
	\end{subfigure}
	\begin{subfigure}[t]{0.32\linewidth}
		\includegraphics[width=0.8\linewidth]{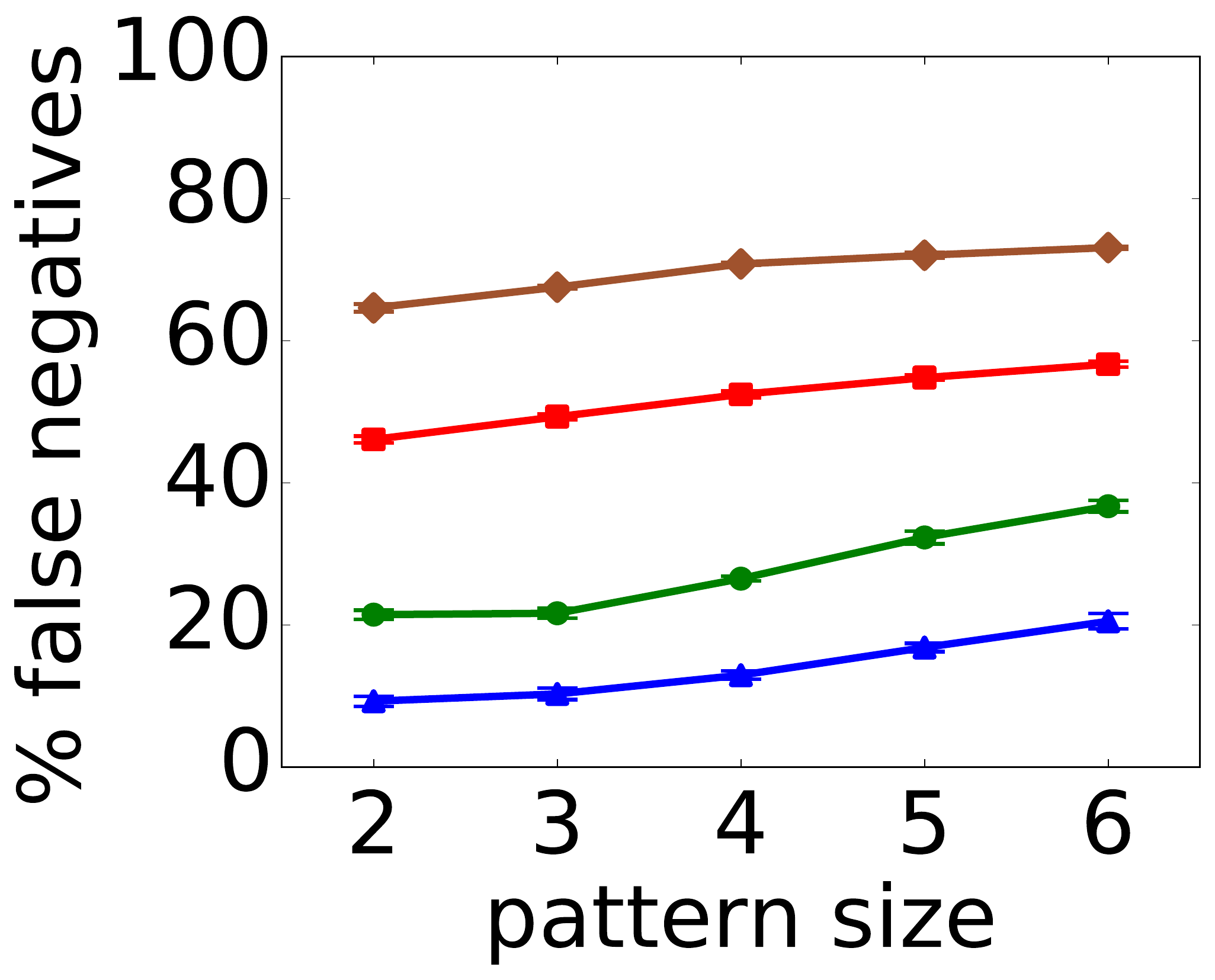}
		\caption[]{Q1: First selection policy}
		\label{fig:q1-fn-first}
	\end{subfigure}
	\begin{subfigure}[t]{0.32\linewidth}
		\includegraphics[width=0.8\linewidth]{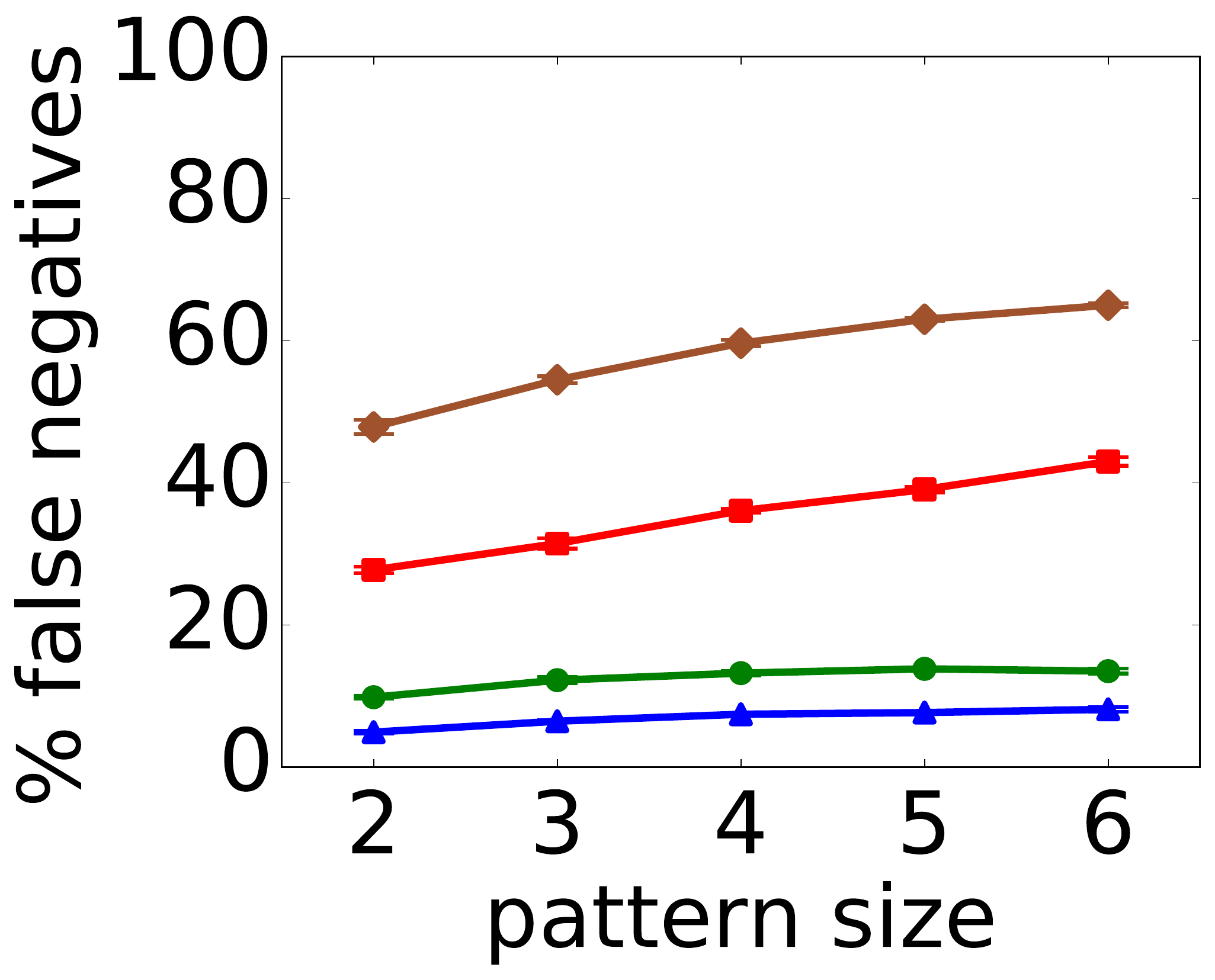}
		\caption[]{Q1: Last selection policy}
		\label{fig:q1-fn-last}
	\end{subfigure}
	\begin{subfigure}[t]{0.32\linewidth}
		\includegraphics [width=0.8\linewidth]{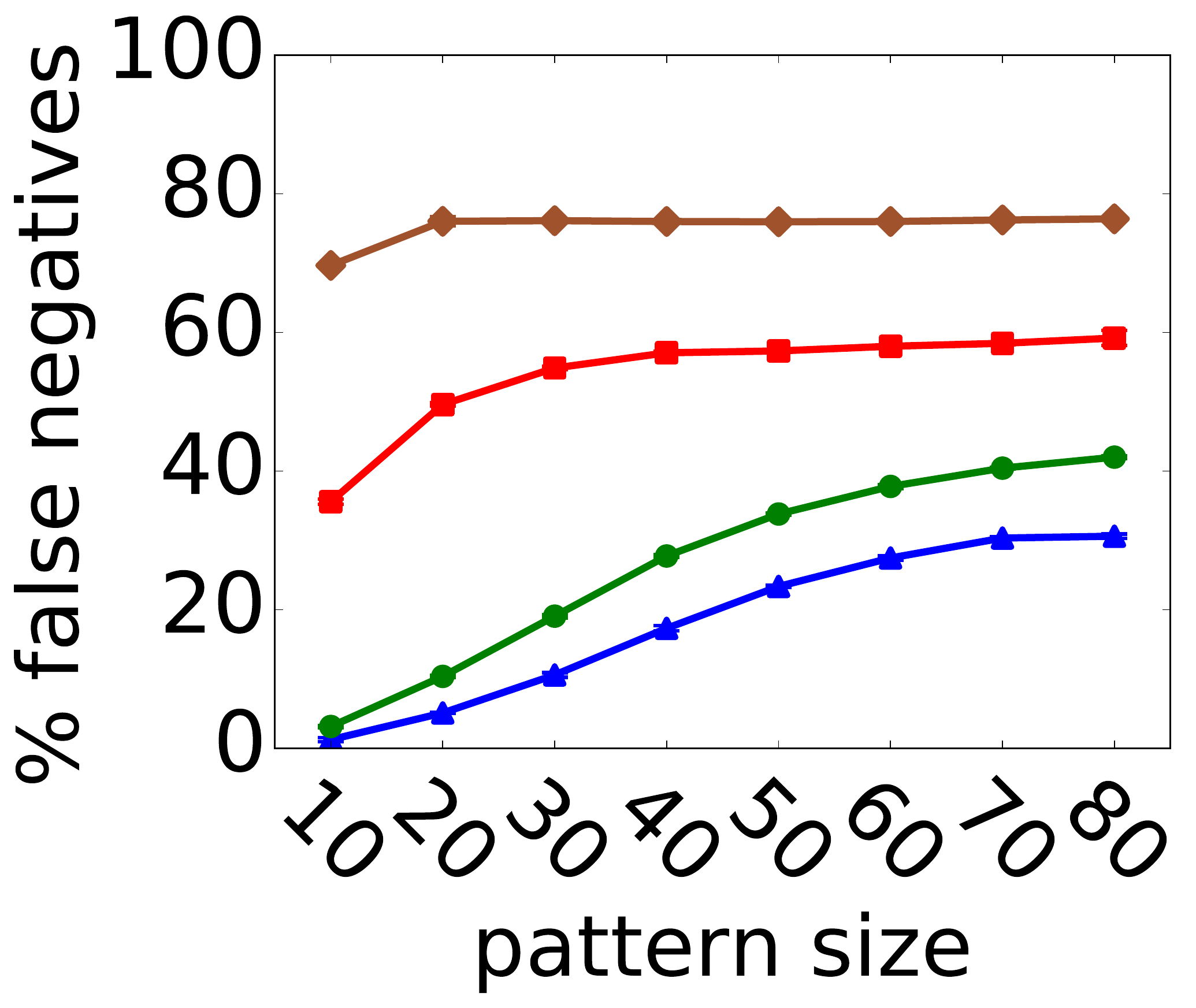}
		\caption[]{Q2: First selection policy}
		\label{fig:q2-fn-first}
	\end{subfigure}
	\begin{subfigure}[t]{0.32\linewidth}
		\includegraphics [width=0.8\linewidth]{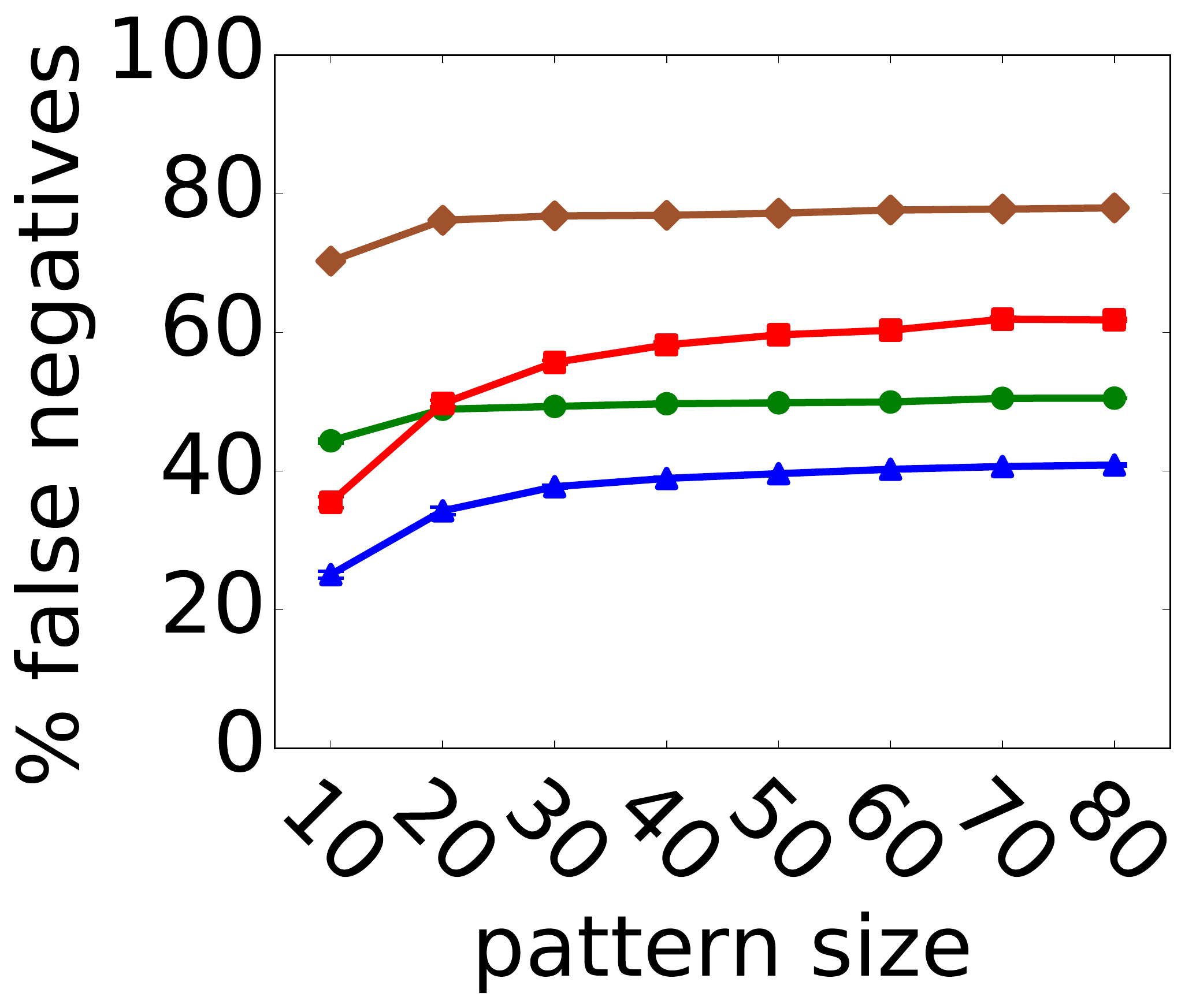}
		\caption[]{Q2: Last selection policy}
		\label{fig:q2-fn-last}
	\end{subfigure}
	\begin{subfigure}[t]{0.32\linewidth}
		\includegraphics [width=0.8\linewidth]{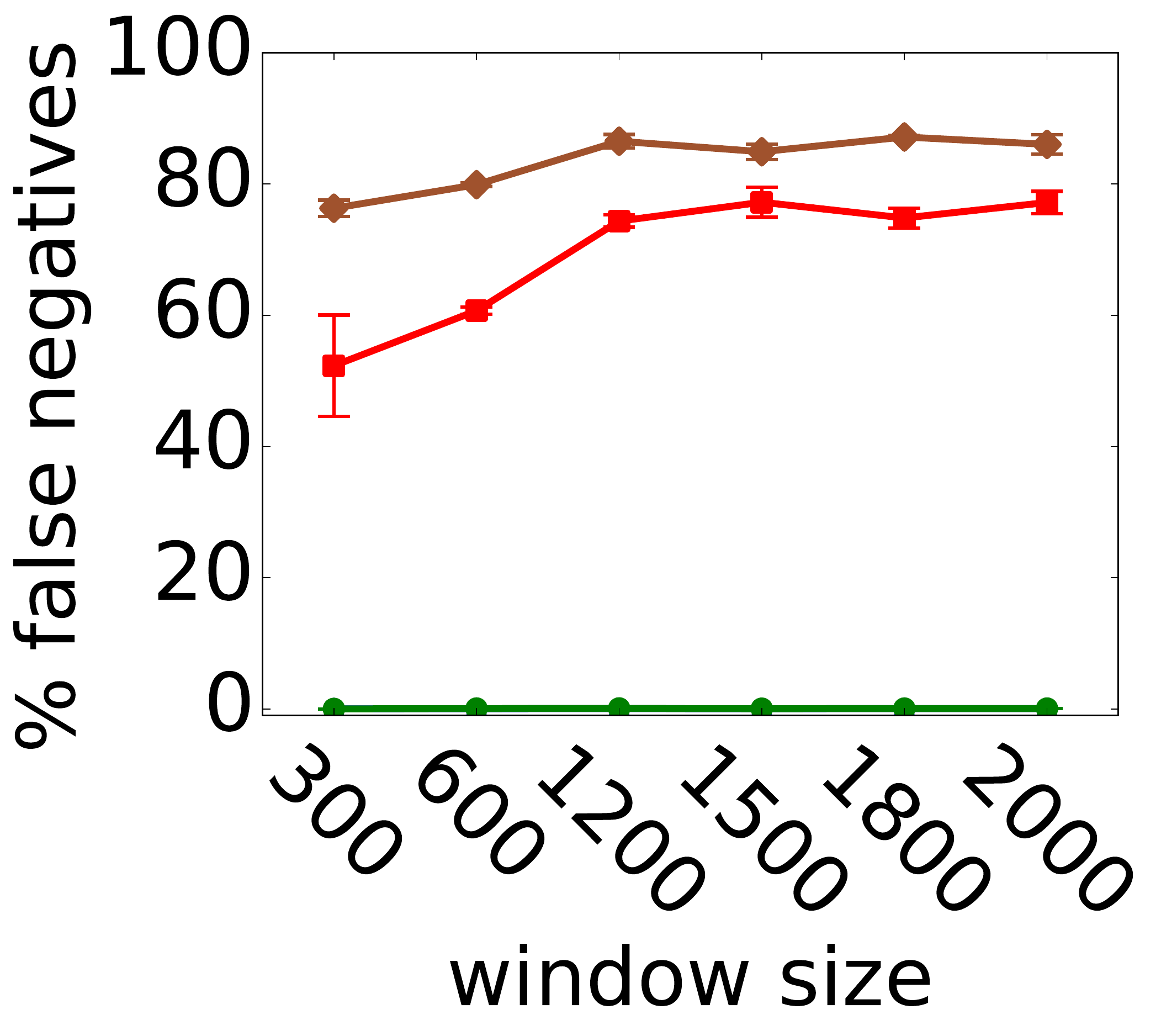}
		\caption[]{Q3: First selection policy}
		\label{fig:q3-fn-first}
	\end{subfigure}
	\begin{subfigure}[t]{0.32\linewidth}
		\includegraphics [width=0.8\linewidth]{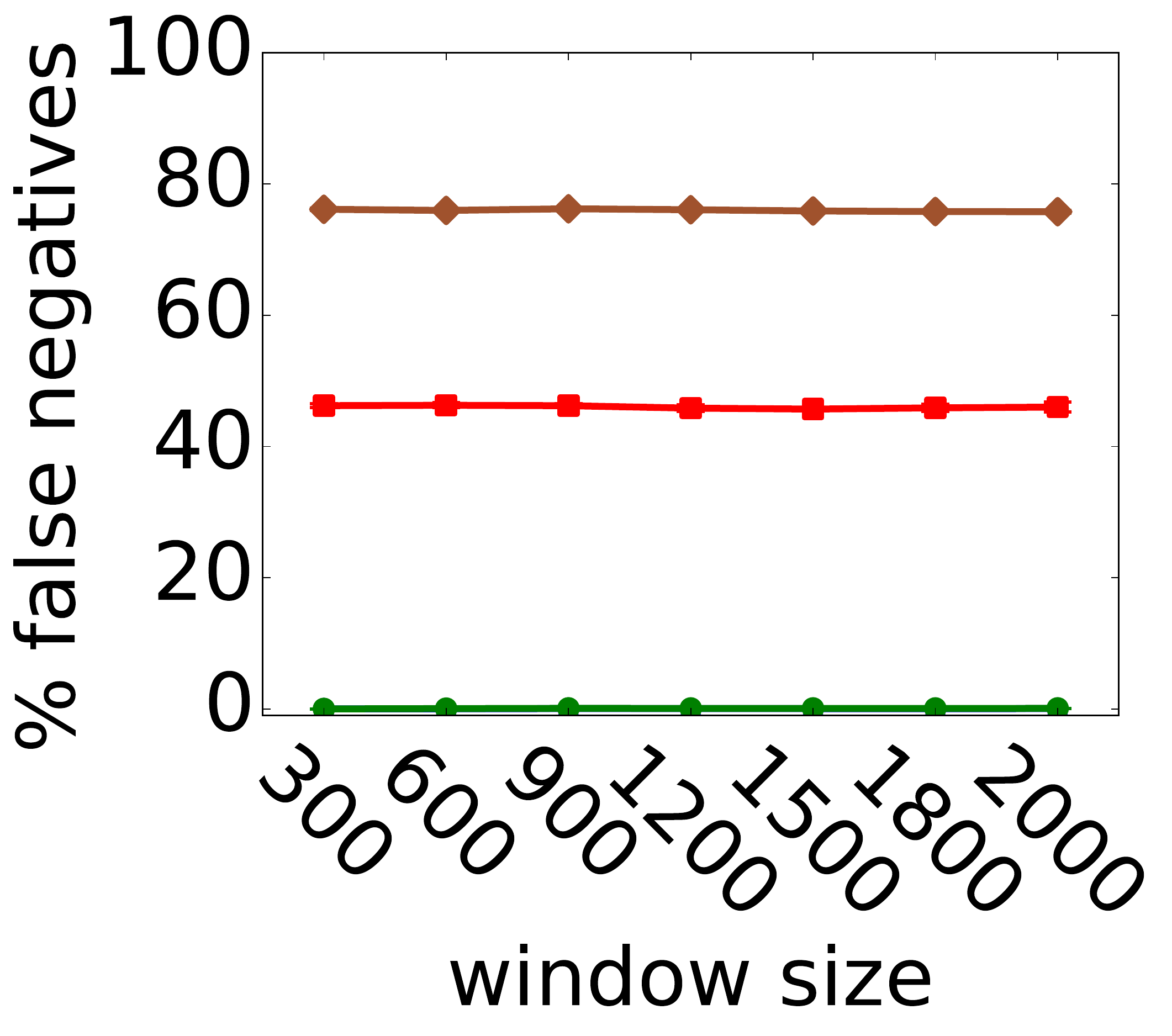}
		\caption[]{Q4: First selection policy}
		\label{fig:q4-fn-first}
	\end{subfigure}
	\vspace{-0.4cm}
	\caption{False negatives for Q1, Q2, Q3 and Q4 with input event rates $R1$ and $R2$.}
	\label{fig:quality_fn}
	\vspace{-0.4cm}
\end{figure*}

\paragraph{Number of false negatives.} 
Figures \ref{fig:q1-fn-first} and \ref{fig:q1-fn-last} show results for Q1 with the first and last selection policy, respectively, while Figures \ref{fig:q2-fn-first} and \ref{fig:q2-fn-last} show results for Q2 with the two selection policies.
Similarly, Figures \ref{fig:q3-fn-first} and \ref{fig:q4-fn-first} show results for Q3 and Q4 using the first selection policy. 
The x-axis in Figures \ref{fig:q1-fn-first}, \ref{fig:q1-fn-last}, \ref{fig:q2-fn-first} and \ref{fig:q2-fn-last} represents the pattern size, while in Figures \ref{fig:q3-fn-first} and \ref{fig:q4-fn-first}, the x-axis represents the window size. In all figures, the y-axis represents the percentage of false negatives. Moreover, all figures show the result of applying \framework~and BL while  using the two input event rates $R1$ and $R2$.

Using a large ratio of pattern size to window size  may result in making the pattern more sensitive to any event drop. Thus, dropping even a few events may hinder detecting the pattern and hence it may result in more false negatives depending on the query. This can be observed in the following results.
In Figure \ref{fig:q1-fn-first}, using the input event rate $R1$ and first selection policy, the percentage of false negative as a result of using \framework~(cf. Figure \ref{fig:q1-fn-first} , the blue line) is 9\% when the pattern size is 2 and it increases with the increment in the pattern size, where it reaches 21.2\% with a pattern of size 6. Similarly, the percentage of false negatives using BL increases from 45.6\% to 55.9\% with the pattern sizes 2 and 6 (cf. Figure \ref{fig:q1-fn-first} , the red line). Also, the percentage of false negatives increases with the higher input event rate $R2$ for both \framework~and BL (cf. Figure \ref{fig:q1-fn-first} , the green and brown lines, respectively). This is because higher input rate implies more dropping.  However, it shows a similar behavior with the increase in the pattern size as in $R1$. In Figure \ref{fig:q1-fn-first}, \framework~produces better results than BL in all cases where the performance of \framework~is higher than the performance of BL by up to 5 and 3.2 times using the event input rates $R1$ and $R2$, respectively.

Q1, using last selection policy, shows similar behavior to the first selection policy behavior as depicted in Figure \ref{fig:q1-fn-last}. The percentage of false negatives for \framework~and the input event rate $R1$ increases from 4\% to 6\% for increasing pattern sizes of 2 to 6. Meanwhile, the percentage of false negatives for BL increases from 27\% to 42\% for the same pattern sizes of 2 to 6 and with the same input rate $R1$. This shows that \framework~performs better than BL by up to 7 times. Similar results are shown when using the input event rate $R2$.

The results of Q2 has similar behavior to the results of Q1.
Figure \ref{fig:q2-fn-first} shows that \framework~performs better than BL by up to 30 and 24 times using the input event rates $R1$ and $R2$, respectively, while using the first selection policy. The performance of \framework~while using last selection policy is very similar to that of the first selection policy, where it significantly outperforms BL for both input event rates $R1$ and $R2$.

Figures \ref{fig:q3-fn-first} and \ref{fig:q4-fn-first}, i.e., running Q3 and Q4 with the first selection policy, show that the performance of \framework~is much better using sequence operator that matches an exact pattern, where the percentage of false negatives is almost zero with both input event rates $R1$ and $R2$. Using the \textit{sequence} operator ensures that every time only the \textit{exact same event types} would match the pattern and construct the complex events. This results in higher utility values for those event types and the shedding is extremely accurate. On the other hand, the \textit{sequence with any} operator allows any event type to match the pattern and hence it results in distributing the utility values more sparsely between different available event types in windows. This is the reason why \framework~performs exceedingly well for sequence operator. 
As a result, the performance of \framework~for Q3 and Q4 is enormously better than the performance of BL which produces a high percentage of false negatives. 
Moreover, our results clearly show that the presence of repetitions in the sequence operator does not impact the performance of \framework~(cf. \ref{fig:q3-fn-first} and \ref{fig:q4-fn-first}). 

\begin{figure}[t]
	\centering
	\begin{subfigure}[t]{0.99\linewidth}
		\centering
		\includegraphics[width=0.8\linewidth]{figures/results/quality/soccer/legend}
	\end{subfigure}
	\hfil%
	\begin{subfigure}[b]{0.49\linewidth}
		\includegraphics[width=0.95\linewidth]{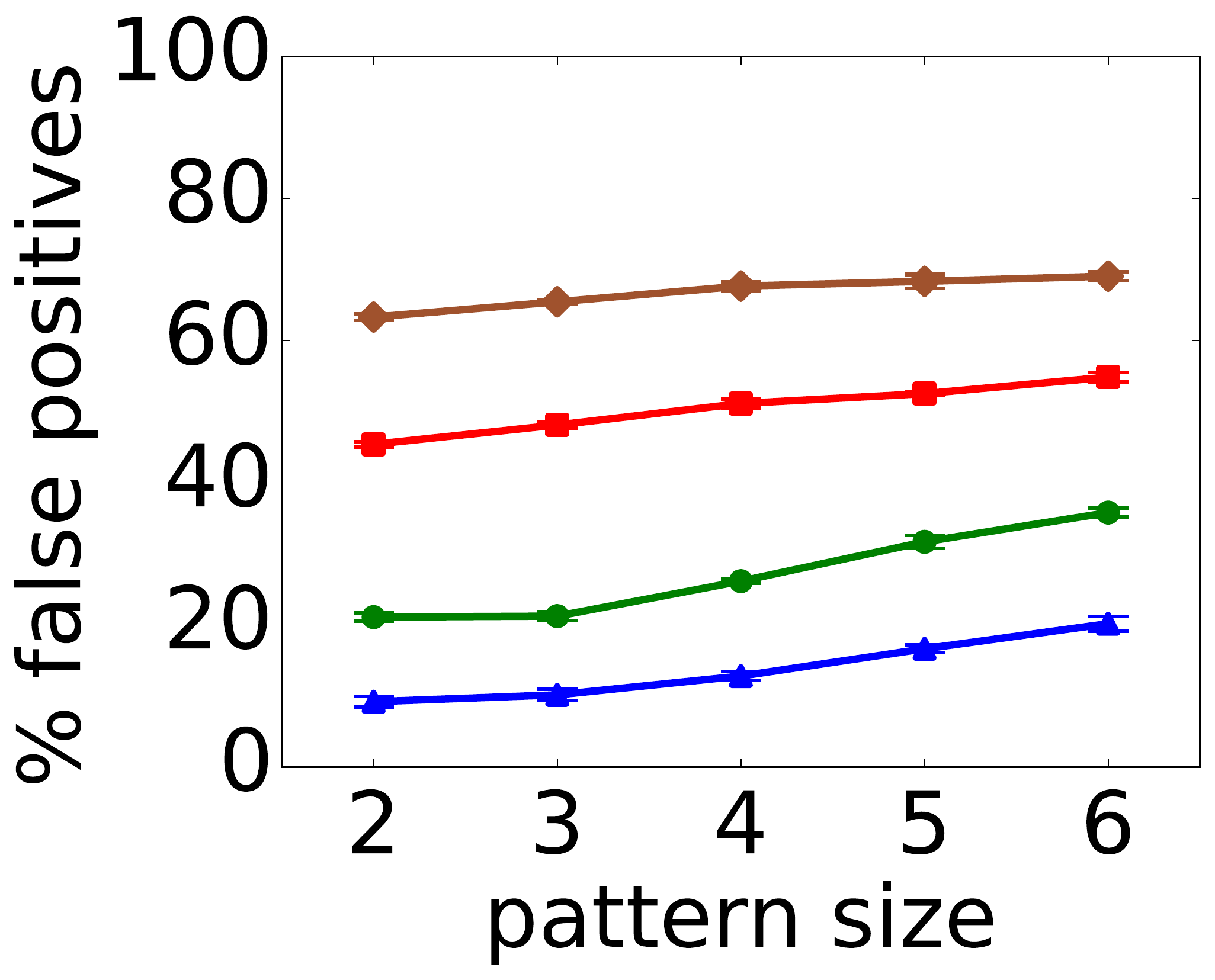}
		\vspace{-0.25cm}
		\caption{Q1: First selection policy}
		\label{fig:q1-fp-first}	
	\end{subfigure}
	\hfill	%
	\begin{subfigure}[t]{0.49\linewidth}
		\includegraphics[width=0.95\linewidth]{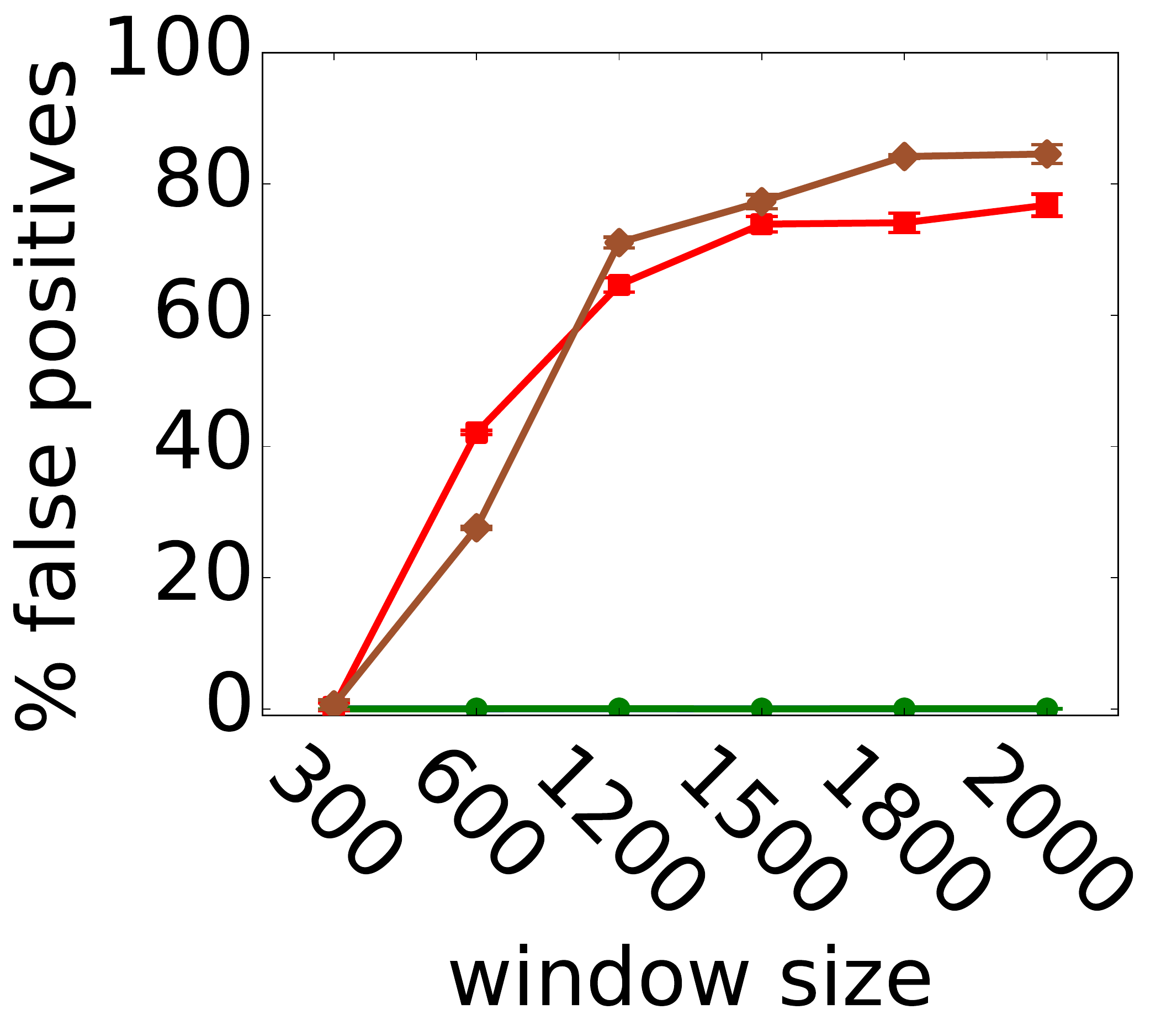}
		\vspace{-0.25cm}
		\caption{Q3: First selection policy}
		\label{fig:q3-fp-first}	
	\end{subfigure}
	\vspace{-0.4cm}
	\caption{False positives for Q1 and Q3 with  $R1$ and $R2$.}
	\label{fig:quality_fp}	
	\vspace{-0.4cm}
\end{figure}

\paragraph{Number of false positives.} 
Figures \ref{fig:q1-fp-first} and \ref{fig:q3-fp-first} show results for Q1 and Q3, respectively,  with the first selection policy. We do not show Q2, Q4, and the last selection policy as they have similar results.  In Figure \ref{fig:q1-fp-first}, the x-axis represents the pattern size while the x-axis in Figure \ref{fig:q3-fp-first} represents the window size. In both Figures, the y-axis represents the percentage of false positives. Moreover, both figures show the results of applying \framework~and BL while using the two input event rates $R1$ and $R2$. 

The percentage of false positives in Figure \ref{fig:q1-fp-first} shows a similar behavior to the percentage of false negatives in Figure \ref{fig:q1-fn-first}, i.e., Q1 with the first selection policy. The percentage of false positives increases with the increment in the pattern size, i.e., the increment in ratio of pattern size to window size. This is because, the probability to drop a primitive event that contributes to detect a complex event in a window increases with a higher pattern size, thus increasing the  percentage of false negatives as we showed above. Moreover, since Q1 uses the \textit{sequence with any} operator, the probability to find any other primitive event, i.e., any other defender event, alternative to the dropped primitive event is high, which can result falsely in detecting a new complex event, i.e.,  a false positive. As in false negative experiment, the percentage of false positives increases with the input event rate. 
\framework~performs better than BL up to 4.8 and 3.2 times using the event input rates $R1$ and $R2$, respectively.

The percentage of false positives in Figure \ref{fig:q3-fp-first} also shows a similar behavior to  the percentage of false negatives in Figure \ref{fig:q3-fn-first}, i.e., Q3 with the first selection policy, where the percentage of false positives is almost zero for \framework. However,  the percentage of false positives for BL increases with the increment in the window size. 
This is because, Q3 uses the \textit{sequence} operator which matches an exact pattern. Thus, if a primitive event that contributes to a complex event in a window is dropped, it is hard to find an alternative primitive event to the dropped primitive event in smaller windows. The probability to find this alternative primitive event in a window increases with the increment in the window size, thus increaing the number of falsely detected events, i.e., increasing the percentage of false positives.

\begin{figure}[t]
	\includegraphics[width=0.50\linewidth]{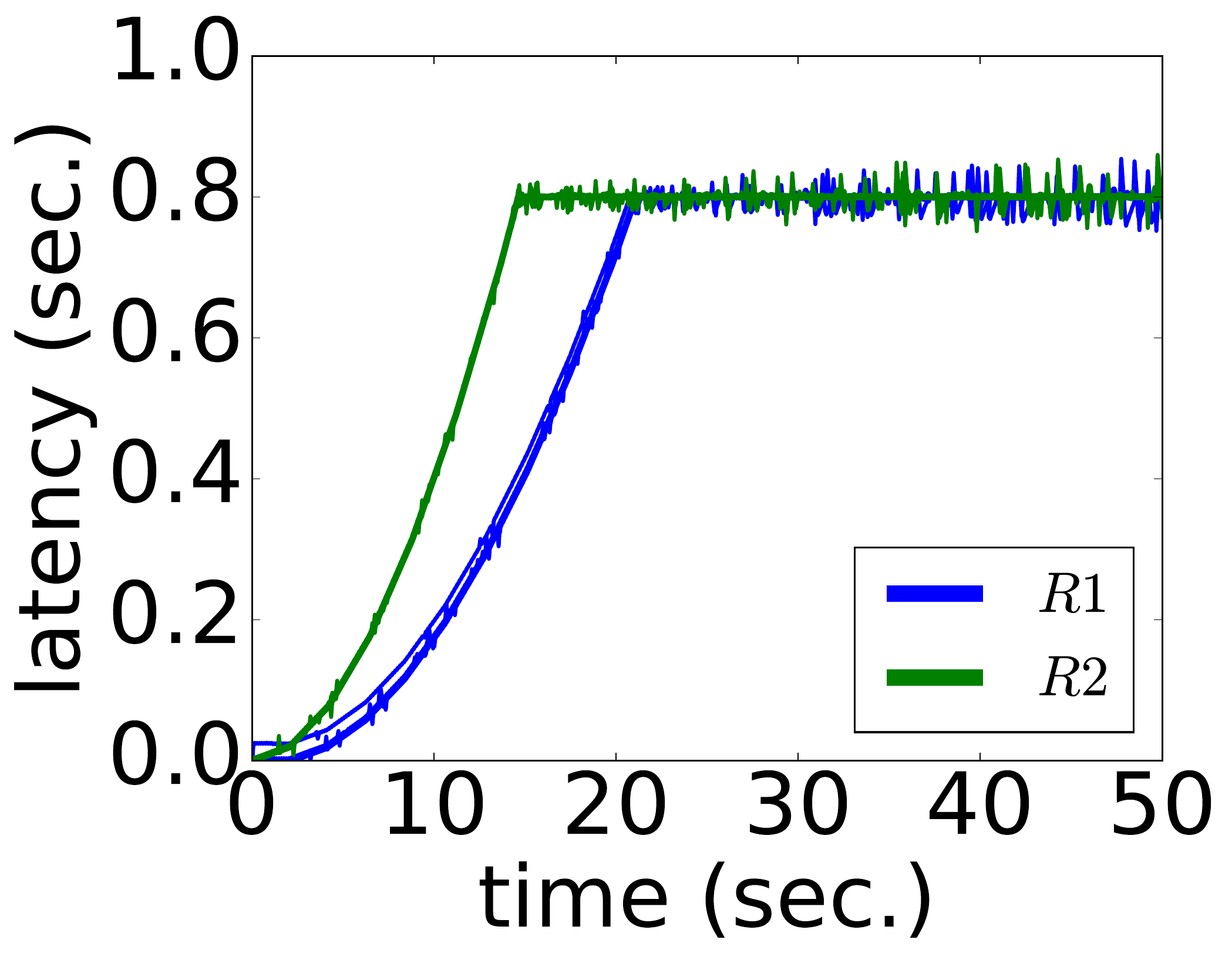}
	\caption{Event processing latency.}
	\label{fig:q1-event-latency}
	\vspace{-0.6cm}
\end{figure}

\paragraph{Maintaining the given latency bound.}
 The main goal of \framework~is to maintain the given latency bound. Hence, here, we discuss the ability of \framework~in keeping the given latency bound. Figure \ref{fig:q1-event-latency} shows the incurred event latency ($l_e$) in case of running Q1 with the input event rates $R1$ and $R2$. The results of other queries show similar behavior and hence they are not shown.  The figure shows that \framework~never violated the given latency bound ($LB = 1$ second) and it always keeps the event latency around ($f*LB$) which is 800 milliseconds in this experiment.

\textbf{\textit{Impact of variable window size on the quality of results.}}
Now, we show the impact of variable window size on the quality of results. Using time-based or pattern-based sliding window may result in splitting the incoming event stream into windows of different sizes. However, $UT$ has a fixed number of positions/events $N$, where $N$ represents the average window size, given $bs= 1$. Hence, we must map the incoming windows of different sizes to $N$ as we showed above in Section \ref{sec: variable-windows}.

The ideal window size should be $N$, however, in case the incoming windows are larger or smaller than $N$, the quality of results might degrade because of the variations in relative positions of events in windows . 
To evaluate that, we run experiments with Q1 and Q2 where we use several window sizes during model building to enforce having different number of events per window. 

For Q1, we use a pattern size of $n = 5$. Moreover, we use a time-based window of the following sizes: $ws$= 12 , 14, 16 ,18 and 20 seconds. The average seen window size is $\approx$ 800 events and hence we use $N =800$ to build $UT$.  As the window size $ws = 16$ seconds contains around 800 events ($\approx N$), we use it as a reference window size in our results and refer to it as a window of size 100\%. We represent the window sizes as percentage values compared to the  reference window size (i.e., $ws= 16$ seconds) and hence the used windows are of the following sizes: 75\%, 87\%, 100\%, 112\% and 125\%. 

For Q2, we use a pattern size of $n = 20$. Again, we use a time-based window of the following sizes: $ws$= 180 , 200, 240 , 260 and 300 seconds.   The average seen window size is $\approx$ 2000 events and hence we use $N =2000$ to build $UT$. The window size $ws = 240$ seconds contains around 2000 events ($\approx N$) therefore we use it as a reference window size in our results and refer to it as window of size 100\%. Again, we represent the window sizes as a percentage value compared to $ws= 240$ seconds and hence the used windows are of the following sizes: 75\%, 83\%, 100\%, 108\% and 125\%.

For  both queries, during the model building, we change the window size between the above given window sizes randomly to ensure that our model has learned from several window sizes and not only from one window size.
During load shedding, we use one of the window sizes of the above given window sizes to check the impact of this window size on the quality of results.

\begin{figure}[t]
	\centering
	\begin{subfigure}[t]{0.49\linewidth}
		\includegraphics[width=0.95\linewidth]{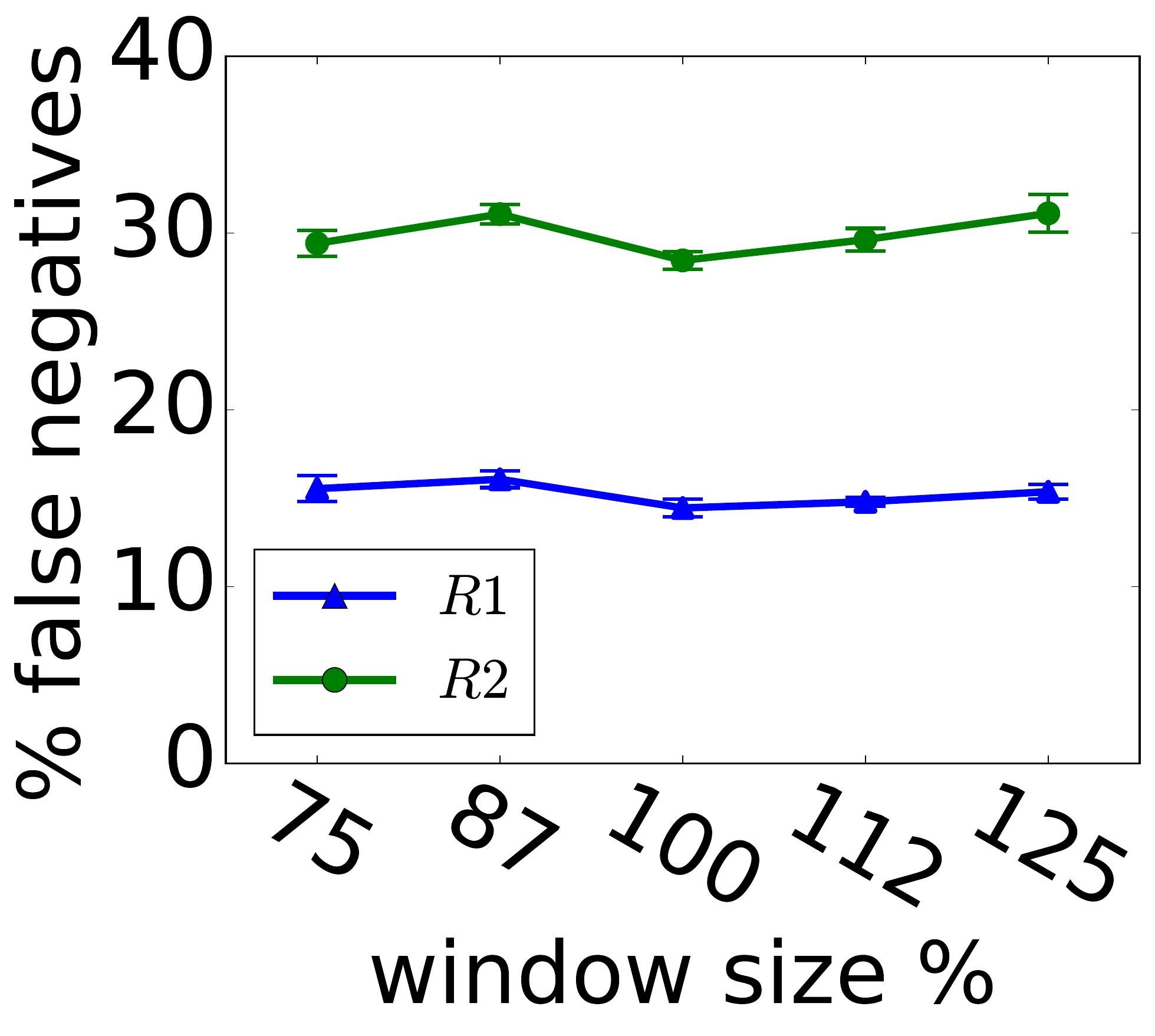}
		\caption{Q1: First selection policy}
		\label{fig:q1:variable-window}
	\end{subfigure}%
	\hfill%
	\begin{subfigure}[t]{0.49\linewidth}
		\includegraphics[width=0.95\linewidth]{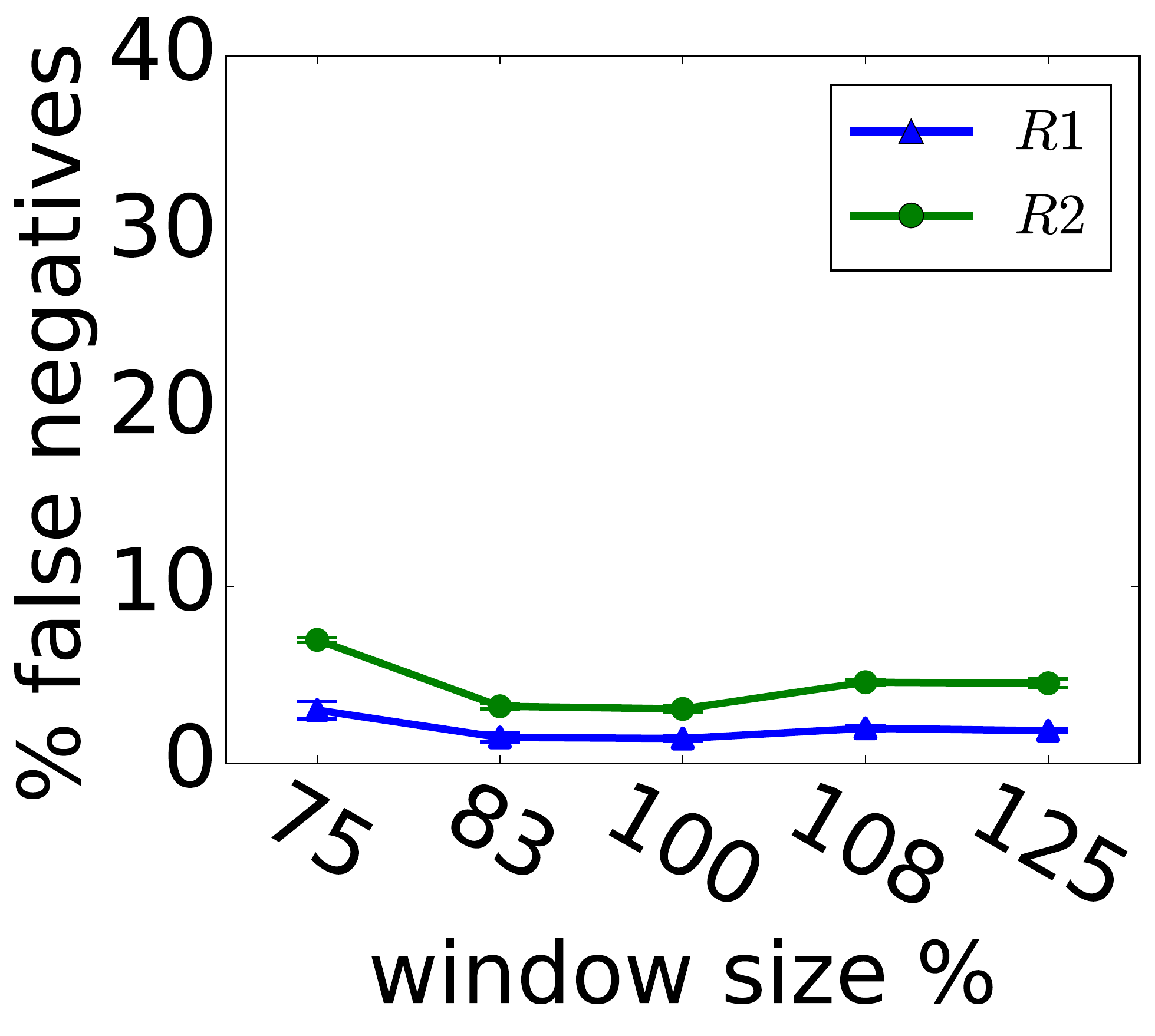}
		\caption{Q2: First selection policy}
		\label{fig:q2:variable-window}
	\end{subfigure}%
	\vspace{-0.4cm}		
	\caption{ Impact of variable window size on the quality.}
	\label{fig:variable-window}
	\vspace{-0.2cm}
\end{figure}

Figure \ref {fig:variable-window} depicts the percentage of false negatives for both Q1 and Q2. The x-axis represents the percentage of  window size compared to the reference window size and the y-axis represents the percentage of false negatives. 
Figure \ref{fig:q1:variable-window} shows results for Q1 with the two event input rates $R1$ and $R2$, respectively while Figure \ref{fig:q2:variable-window} shows results for Q2 with the two rates, all using the first selection policy. We observed similar results for Q1 and Q2 using the last selection policy and also for the number of false positives experiments and hence we don't show them.

Figure \ref{fig:q1:variable-window} shows that the percentage of false negatives is only slightly influenced by the used window size with both input event rates $R1$ and $R2$. Hence, more than one event in a window can be mapped to a single position in $UT$ in case $ws > N$ or one event in a window can be mapped to several positions in $UT$  when $ws < N$ without having a considerable impact on the number of false negatives.

The impact of window size is clearer in Figure \ref{fig:q2:variable-window}, i.e., in Q2. 
In this figure, the percentage of false negatives increases when the difference between $N$ and the window size increases.  
The reason behind this is that Q2 has a longer pattern size than Q1 which makes it more sensitive to the relative event positions in windows. Moreover, the number of event types (i.e., MLE) that start a new match in Q2 is higher than the number of event types that start a new match in Q1 (only two strikers). 
In Q2, each event in MLE may impact different stock companies and hence the utility values are more distributed between different event types and over the whole window. Meanwhile, in Q1, the distribution of the utilities is more focused on specific number of defenders. 

\textbf{\textit{Impact of bin size on the quality of results.}} 
A big bin size might degrade the quality of results since it reduces the accuracy in $UT$ of the important positions in the incoming windows. To analyze the impact of bin size on the quality of results, we run experiments with Q1 and Q2. We use a pattern of size $n= 5$ and $n =20$ and a window of size $ws= 15$ seconds and $ws = 240$ seconds for Q1 and Q2, respectively. In addition, we use the following bin sizes for both queries: $bs= $ 1, 2, 4, 8, 16, 32, 64.
 
\begin{figure}[t]
	\centering%
	\begin{subfigure}[t]{0.49\linewidth}
		\includegraphics[width=0.95\linewidth]{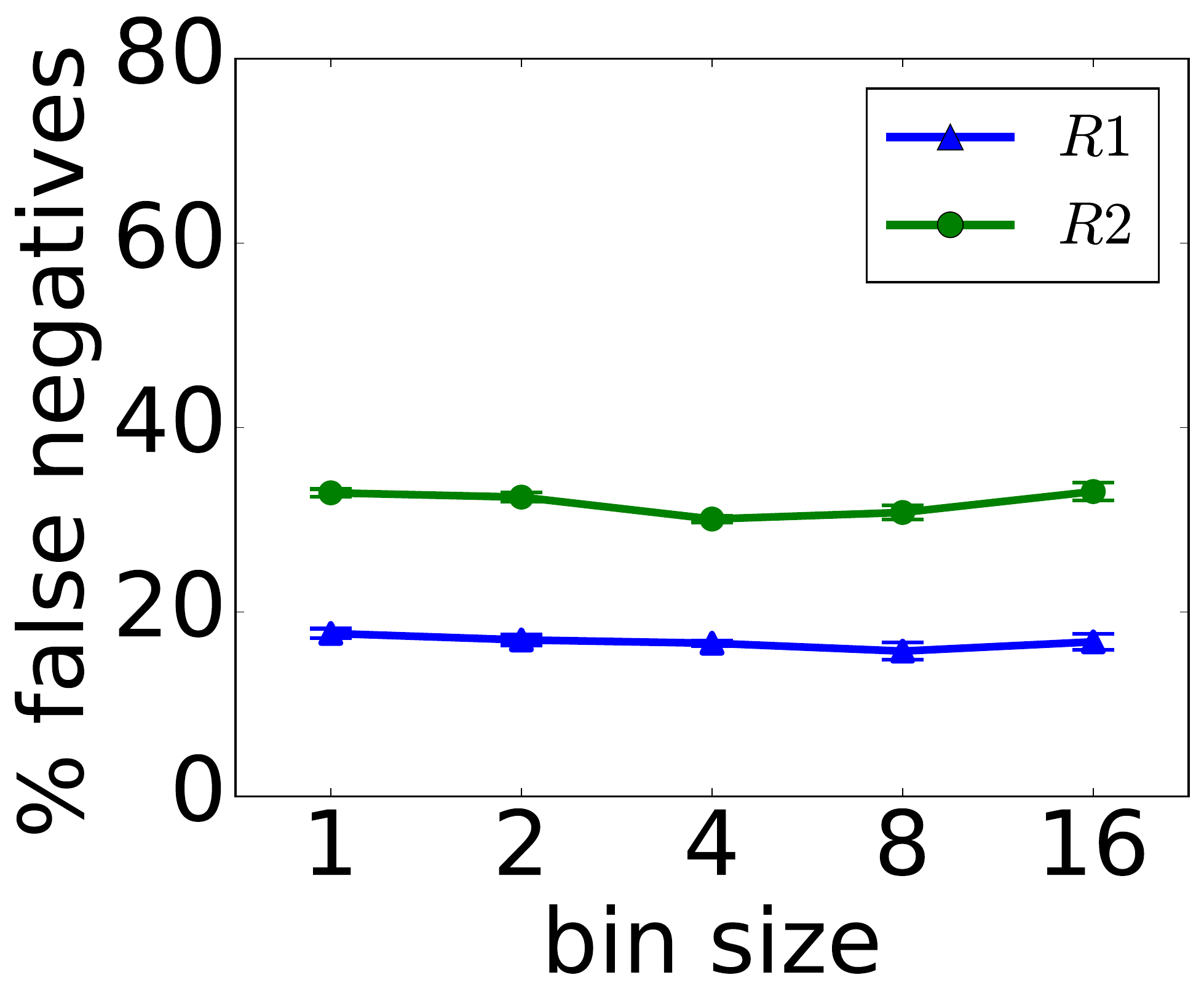}
		\caption{Q1: First selection policy}
		\label{fig:q1:bin-size}
	\end{subfigure}%
	\hfill%
	\begin{subfigure}[t]{0.49\linewidth}
		\includegraphics[width=0.95\linewidth]{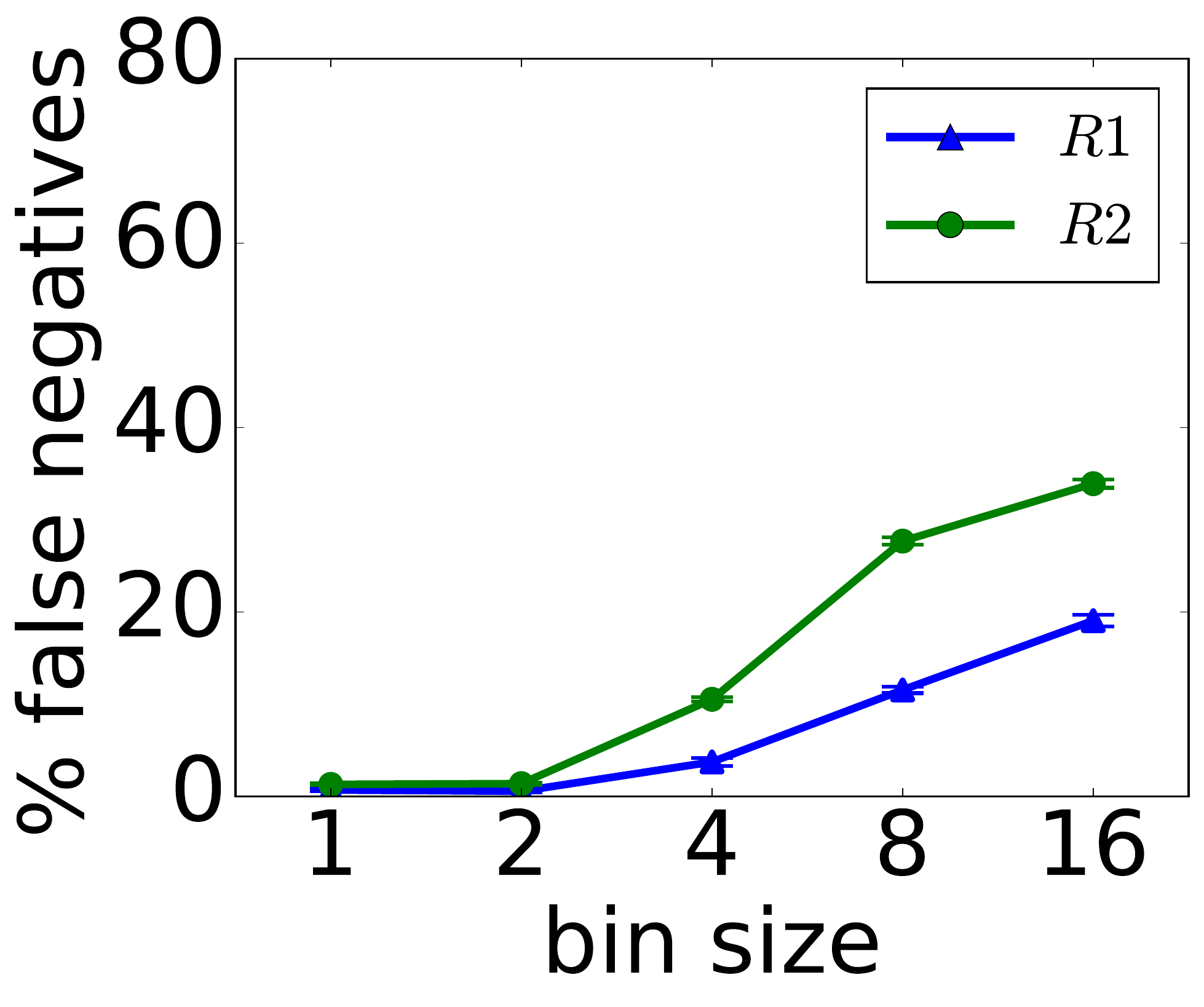}
		\caption{Q2: First selection policy}
		\label{fig:q2:bin-size}
	\end{subfigure}%
	\vspace{-0.4cm}
	\caption{ Impact of bin size on the quality.}
	\label{fig:bin-size}
	\vspace{-0.6cm}
\end{figure}

Figure \ref {fig:bin-size} depicts the percentage of false negatives for both queries with the first selection policy. The x-axis represents the bin size and the y-axis represents the percentage of false negatives. 
We observed similar results for Q1 and Q2 using the last selection policy and also for the number of false positives experiments and hence we don't show them.
 
Figure \ref{fig:q1:bin-size} depicts results for Q1 with the rates $R1$ and $R2$ where it shows that the percentage of false negatives is slightly influenced by the used bin size for both input event rates $R1$ and $R2$.

Figure \ref{fig:q2:bin-size} depicts results for Q2 with the rates $R1$ and $R2$ where it shows that the percentage of false negatives increases with the used bin size. The reason here is again similar to the reason in the variable window size experiment.

\textbf{\textit{Run-time overhead of the LS.}}
Load shedding is used in systems that already face overload and hence the LS overhead must be considerably small compared to the event processing overhead. Our LS performs only a single lookup in the utility table $UT$ to decide whether or not to drop an event from a window and hence its time-complexity is O(1). Thus, it is a lightweight load shedding strategy.

An important parameter that impacts the LS overhead is the window size.  A large window may not fit in the system caches and  cost higher lookup time and hence higher overhead. 

To show the overhead of the LS, we run experiments for Q2 with the two input event rate $R1$ and $R2$ and use a window of the following sizes: $ws= $ 240, 360, 480, 960, 1920 seconds, where the approximate window sizes in events are 2000, 3000, 4000, 8000 and 16000 events, respectively. We used these approximate window sizes in events as a dimension for  $UT$, i.e., $N= ws$.  We observed similar behavior for other queries and hence we do not show them.

\begin{figure}[t]
	\centering
	\includegraphics[width=0.50\linewidth]{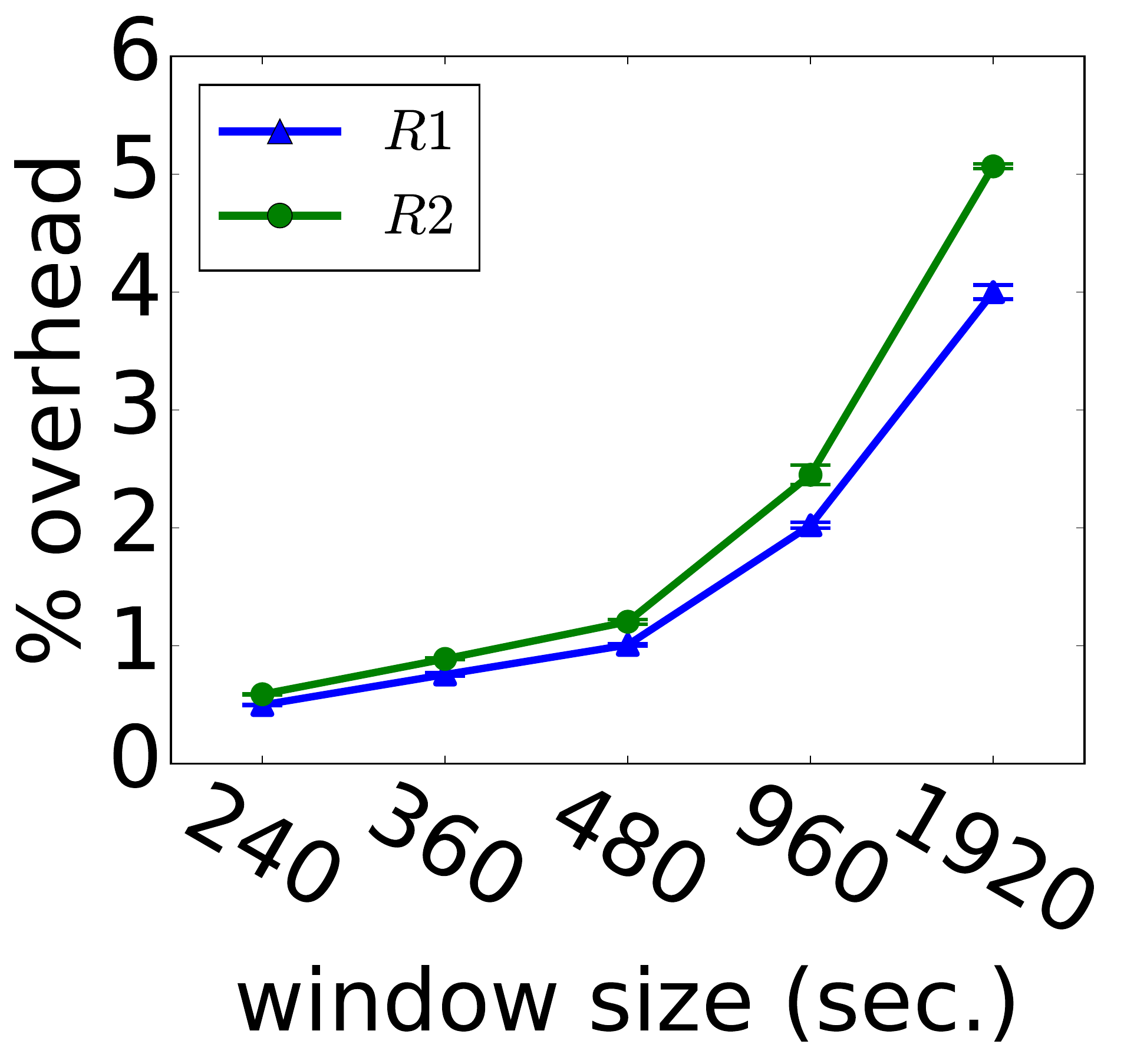}
	\vspace{-0.3cm}
	\caption{ Q2: Overhead of the LS.}
	\label{fig:q2:overhead}
	\vspace{-0.6cm}
\end{figure}

Figure \ref{fig:q2:overhead} depicts the overhead of the LS for Q2. The x-axis represents the used window size and the y-axis represents the percentage time the LS needs compared to the actual event processing time. As expected, the overhead of our LS increases  with the used window size. In the figure, the overhead increases from less than 1\% with the window of size 240 seconds ($\approx 2000$ events) to $\approx$  5\% with the window of size 1960 seconds ($\approx 16000$ events).  Please note that the result shown for the window size 1960  seconds is with respect to a large table size where $M= 500$ (number of stock companies) and $N= 16000$.
However, the overhead is still small compared to the actual event processing time.  Hence, our load shedding strategy can maintain the given latency bound with a low overhead. Moreover, the overhead of the window size can be reduced by increasing the the bin size ($bs$). In addition, improving the utility table locality in the memory can further reduce the overhead of LS.

\textbf{\textit{Results Discussion.}}
\framework~performs much better than BL for all queries, datasets and selection policies. 
However, the performance of \framework~varies for different classes of operators.
The performance of \framework~is exceptionally good for the  \textit{sequence} and \textit{sequence with repetition} operator with a negligible number of false negatives and positives.
The \textit{sequence} operator ensures  that every time only the same event types would match the pattern and this results in higher utility values for those event types. 
On the other hand, the \textit{any} operator matches any event regardless of its type. Hence, the utility of the events are more sparse which adversely impacts the performance of \framework. 
Further, \framework~shows its robustness against variable window size and bin size, where the quality of results is only slightly influenced by a window size which is different from $N$ or by a higher bin size. Moreover, the overhead of LS component in \framework~is very low compared to the actual processing overhead which makes \framework~suitable for real-time complex event processing. 

\vspace{-0.2cm}
\section{Related Work}
\label{sec:related_work}

Complex event processing has emerged as an important paradigm to process event streams on-the-fly. Different event definition languages~\cite{1:ch1994snoop,1:cu2010tesla, Wu:2006:HCE:1142473.1142520}  have specified the rules for CEP systems. To precisely match the events, ~\cite{1:ch1994snoop,1:cu2010tesla, 1:Zimmer:1999:SCE:846218.847253} define selection and consumption policies.
CEP systems are used to process huge event streams and required to provide a high performance. Hence, the CEP operator graph is usually distributed on several compute nodes and each operator might be parallelized to speedup its processing~\cite{1:apache_storm, 1:s4:5693297}. A powerful parallelization technique is data-parallel CEP ~\cite{ 1:Balkesen:2013:RRI:2488222.2488257, CastroFernandez:2013:ISO:2463676.2465282, 1:flux:1260779, 1:s4:5693297, 1:apache_storm}.
Data-parallel CEP is mainly divided in two categories, namely window-based parallelization~\cite{1:Balkesen:2013:RRI:2488222.2488257, 1:apache_storm} and key-based parallelization~\cite{CastroFernandez:2013:ISO:2463676.2465282, 1:flux:1260779, 1:s4:5693297, 1:apache_storm}.
The key-based parallelization is limited only to the applications where events could have a key.  


Various approximation techniques are frequently used to avoid resource constraints in various domains such as in-network processing~\cite{Tariq:pleroma, Bhowmik:Addressing}, distributed graph processing~\cite{shang2014auto}, stream processing~\cite{3:Tatbul:2003:LSD:1315451.1315479, 3:Rivetti:2016:LSS:2933267.2933311, 3:Olston:2003:AFC:872757.872825}, etc.
In fact, load shedding has been proposed by several research groups \cite{3:Tatbul:2003:LSD:1315451.1315479, 3:Tatbul:2006:WLS:1182635.1164196, 3:Rivetti:2016:LSS:2933267.2933311, 3:Olston:2003:AFC:872757.872825, 3:Kalyvianaki:2016:TFF:2882903.2882943, Manku:2002:AFC:1287369.1287400, 1320010, 8622265} in the stream processing domain. The idea is to drop events in a way that reduces the system load but still provides the maximum possible quality of result. Hence, the crucial question here is which events (tuples) to drop so the quality of result is not impacted drastically. 
In \cite{3:Tatbul:2003:LSD:1315451.1315479, 3:Olston:2003:AFC:872757.872825}, the authors assumed that the tuples have different utility values-- which reflects their importance and impact on the quality of result. In case of overload, the tuples with low utility values are dropped.
The work in \cite{3:Rivetti:2016:LSS:2933267.2933311} proposes to drop tuples that incur higher processing latency where the authors claim that tuples have the same utility value but may have different processing latency. 
In \cite{Quoc:2017:SAC:3135974.3135989} the authors assume that all tuples have the same utilities and processing latency. They  fairly select tuples for drop from different input streams by combining two techniques, stratified sampling and reservoir sampling.
In contrast to all these works,  in CEP, the utility of events is influenced by other events in the same pattern since  CEP systems perform pattern correlation. Hence, we cannot consider only the utility of each event individually but we have to also take into the consideration other events in the pattern. Moreover, the order of events in patterns and in input event streams is important in CEP (e.g., the  sequence operator) which is not considered in the above works.

The authors in \cite{3:He2014OnLS} have formulated the load shedding problem in CEP as a set of different optimization problems. Two types of load shedding are considered by the authors: integral load shedding in which specific event types or patterns are dropped and fractional load shedding where a uniform sampling is used to keep a portion of event types or pattern matches. However, they do not consider the order of events in patterns and in input event streams which is important in CEP.
In \cite{pSPICE}, the authors proposed a load shedding approach, called pSPICE, to drop partial matches instead of events from a CEP operator's internal state where partial matches that have low probabilities to complete and become complex events are dropped. However, as the authors show, pSPICE is outperformed (w.r.t. quality of results) by event dropping strategies if the partial match completion probability is relatively high.

\vspace{-0.2cm}
\section{Conclusion}
In this paper, we proposed a lightweight load shedding framework, called \framework, for window-based CEP systems which maintains a given latency bound by dropping events while reducing its adverse impact on the quality of results. \framework~uses the type and relative position within windows  of primitive events to predict their utility values and efficiently drops events from incoming windows.
Through extensive evaluations on two real world datasets and a range of popular CEP operators, we show that \framework~outperforms state-of-the-art load shedders for CEP/stream processing systems. \framework~successfully maintains the given latency bound while keeping the degradation in quality of results very low at minimum overhead.  

\balance
\section*{Acknowledgement}
This work was supported by the German Research Foundation (DFG) under the research grant "PRECEPT II" (BH 154/1-2 and RO 1086/19-2).

\bibliographystyle{ACM-Reference-Format}
\bibliography{paper}

\end{document}